\documentclass[10pt]{article}
\usepackage[affil-it]{authblk}
\pdfoutput=1 
\usepackage{fullpage}
\usepackage{url}
\usepackage{amsmath}
\usepackage{amsfonts}
\usepackage{graphicx}
\usepackage{subcaption}
\usepackage{algorithm}
\usepackage{algorithmic}
\usepackage{hyperref}
\usepackage[UKenglish]{babel}
\DeclareMathOperator*{\argmin}{\arg\!\min}

\usepackage{natbib}
\begin{document}
\title{Polyquant CT: direct electron and mass density reconstruction\\ from a single polyenergetic source}

\author{Jonathan H. Mason\textsuperscript{1}%
	\thanks{Electronic address: \texttt{j.mason@ed.ac.uk}; Corresponding author} }
\author{Alessandro Perelli\textsuperscript{1}}\author{William H. Nailon\textsuperscript{1,2}}\author{Mike E. Davies\textsuperscript{1}}
\affil{\textsuperscript{1}Institute for Digital Communications, \\University of Edinburgh, Edinburgh, EH9 3JL, UK}
\affil{\textsuperscript{2}Department of Oncology Physics, Edinburgh Cancer Centre, \\Western General Hospital, Edinburgh, EH4 2XU, UK}

\maketitle
\begin{abstract}
Quantifying material mass and electron density from computed tomography (CT) reconstructions can be highly valuable in certain medical practices, such as radiation therapy planning. However, uniquely parameterising the X-ray attenuation in terms of mass or electron density is an ill-posed problem when a single polyenergetic source is used with a spectrally indiscriminate detector. Existing approaches to single source polyenergetic modelling often impose consistency with a physical model, such as water--bone or photoelectric--Compton decompositions, which will either require detailed prior segmentation or restrictive energy dependencies, and may require further calibration to the quantity of interest. In this work, we introduce a data centric approach to fitting the attenuation with piecewise-linear functions directly to mass or electron density, and present a segmentation-free statistical reconstruction algorithm for exploiting it, with the same order of complexity as other iterative methods. We show how this allows both higher accuracy in attenuation modelling, and demonstrate its superior quantitative imaging, with numerical chest and metal implant data, and validate it with real cone-beam CT measurements.
\end{abstract}
\section{Introduction}
Physically quantifying reconstructions from CT is highly valuable in some medical practices. In radiation therapy for example, the electron density inferred from CT images of the patient allows the dose deposition from the treatment beam to be accurately modelled \citep{Battista1980}. Additionally, quantifying the bone mineral density allows osteoporosis to be characterised and the risk of bone fractures to be assessed \citep{Marshall1996}. Due to the non-linear energy-dependence of X-ray attenuation however, mapping from a set of raw measurements to a consistent physically quantitative reconstruction is not straight forward, and requires both actively accounting for the polyenergetic effects such as beam hardening, and establishing a relation to the quantity of interest.

Mapping from CT to physical density is usually treated in a three step process: linearising the measurements by correcting for scatter, taking the logarithm \citep{Chang2014} and non-linearly calibrating from a polyenergetic to approximate monoenergetic source \citep{Joseph1978}; inverting the linearised projections with analytic or iterative reconstruction algorithms \citep{Fessler2014}; then finally applying a non-linear calibration to mass or electron density \citep{Schneider1996}.

In general, the mapping from a single polyenergetic to a monoenergetic measurement is an ill-posed problem, since the spectral information of the irradiated material is lost with the standard energy integrating detectors \citep{Curry1990}. Whilst imaging the human body however, most tissues may be reasonably modelled with few distinct materials: namely water and bone as in \citep{Joseph1978}. If the amount of each material in a projection is known a priori from a water--bone segmentation, then it is possible to approximately reverse the polyenergetic effects through polynomial fitting or with a look up table. For more accuracy, this may also be brought into the reconstruction model, where \citep{Elbakri2002} demonstrate statistical quantitative imaging of mass density, with the water--bone assumption. Given that the composition is itself highly correlated with density, then the prior segmentation may not be necessary, but estimated during reconstruction \citep{Elbakri2003}.

Another approach is to model the attenuation explicitly in terms of physical processes, given quantitative physical parameters of interest. One such choice is to model the photoelectric and Compton effects in terms of relative atomic number and electron density, which is reasonably accurate for relevant elements and energies \citep{Weber1969,Rutherford1976a,Jackson1981}. Given measurements from two sufficiently different spectra---a technique known as dual-energy CT (DECT)---a projection of Compton attenuation can be uniquely determined \citep{Alvarez1976}, of which electron density is an analytic function \citep{Klein1929}. This DECT technique effectively bypasses the need for assumptions such as water--bone compositions, so should be applicable to a wider range in materials, although one faces practical difficulties in generating the two spectra \citep{Johnson2012}. Additionally, one will expect a loss in accuracy between significantly different elemental compositions, such as soft tissues, bone and metallic implants, since the two parameter model is not consistent over a wide range in atomic species \cite{Jackson1981}.

In \citep{DeMan2001a}, the authors introduce an iterative maximum-likelihood polychromatic algorithm for CT (IMPACT), which models the energy independent factors in the photoelectric--Compton model from \citep{Alvarez1976} as piecewise-linear functions of monoenergetic attenuation, allowing reconstruction from a single source. This method does not require any prior segmentation, and is reported on a wide range of materials including metallic implants. By imposing the energy dependence of the two parameter model however, the physical consistency will also degrade throughout diverse material types, due to the inconsistency of these parameterisations in effective atomic number and electron density \cite{Jackson1981}.

The second conversion from either reconstructed attenuation in Hounsfield units (HU) in \citep{Joseph1978,DeMan2001a} or mass density in \citep{Elbakri2002,Elbakri2003} to electron density is also an ill-posed problem. This is because X-ray interaction depends on the environment of the electrons as well as their density, and will vary considerably with atomic number for imaging energy ranges \citep{Jackson1981,Curry1990}. Although, again given the fact that most human tissues have similar properties, a single piecewise linear fitting is reasonably accurate in practice \citep{Constantinou1992}, though it will not be consistent with synthetic materials such as some plastics \citep{Schneider1996}.

In both the photoelectric--Compton and a material decomposition such as the water--bone model, there is some degree of fitting to materials and energies of interest. Instead, one could use a purely data-centric approach. Here, given a representative set of substances, one could use a model that accurately parameterises the energy dependent attenuation in terms of the quantity of interest, without necessarily any physical justification. One such method is to model the energy dependent attenuation as a piecewise-linear function of quantitative density, which may be fit to a set of materials of interest. When the transitions between linear sections are independent of energy we also get the nice property that the computation order in an iterative method is independent of the number of energies considered.

In this article, we will study in detail the piecewise linear quantitative model for CT, and will show how it may be incorporated in a regularised iterative reconstruction algorithm. Specifically, this provides a generalised method for directly quantifying the electron or mass density of a heterogeneous specimen, and is also able to model hard metallic structures without any prior segmentation. In \citep{Mason2017} we presented a preliminary study using this idea for the specific case of calculating electron density for radiotherapy planning.

In preparation of this manuscript, we became aware of the commercial method DirectDensity\texttrademark\ from Siemens Healthineers\textsuperscript{\textregistered} \citep{Ritter}, which reports direct reconstruction into relative electron density from a single polyenergetic source. This is a preprocessing technique combining bone detection with a projection-based material decomposition similar to \citep{Joseph1978}. In this article we compare against the bone--model of \citep{Elbakri2002}, which has been shown to itself provide superior performance to that of \citep{Joseph1978}.

\subsection{Contributions}
We establish and analyse a general modelling technique to allow direct quantitative reconstruction from a single polyenergetic source, where we study the cases of electron and mass density imaging. Unlike existing approaches of fitting to physical parameterisations of attenuation, such as bone--water or photoelectric--Compton, we fit directly to the data, which we demonstrate is more accurate over a wide range of biological tissues. We show how this model may be incorporated into statistical reconstruction, and propose an algorithm for performing this that allows further convex spatial regularisation to be used. By design, the complexity of using our model does not scale with the number of discrete energies, and will have an order of computational cost $2.5\times$ that of standard monoenergetic iterative algorithms. As another consequence of the fitting constraints, one could also quantify the equivalent attenuation from a mono-energetic source---known as `quasi-monoenergetic' in DECT \cite{Johnson2012}---though this is not evaluated in this work.  We also demonstrate how this model may also directly mitigate metal artefacts, without any need for segmentation.

\subsection{Article Structure}
We begin this article with background material in Section~\ref{sec:background} on X-ray attenuation, existing polyenergetic parameterisations, and the probabilistic measurement model we will invoke for reconstruction. We then propose our generalised data centric model in Section~\ref{sec:eden_atten}, and outline the cases of electron and mass density quantisation for biological tissues, as well as synthetic materials and metal implants. In Section~\ref{sec:quant_recon} we demonstrate how to utilise this model in statistical reconstruction, and detail one such algorithm in Section~\ref{sec:algorithm}. The experimentation in Section~\ref{sec:experiment} evaluates our method with: a model accuracy test in Section~\ref{sec:model_test} against other physical parameterisations; a numerical reconstruction test in Section~\ref{sec:rec_test} with simulated fan-beam CT of a chest and pelvis with metallic hip implants; and reconstruction validation on a real physical phantom scanned with cone-beam CT. We then discuss important considerations in Section~\ref{sec:discuss} leading to conclusions in Section~\ref{sec:conclusions}.

\section{Background} \label{sec:background}
\subsection{The CT measurement model}
In CT, one is able to observe a specimen's attenuation through the radiation intensity after transmission. The magnitude of this is found from the Beer--Lambert law, given for a monoenergetic beam as
\begin{equation}
I_\mathrm{out} = I_\mathrm{in}\exp\left(-\int_{\ell}\mu(\ell)d\ell\right),
\end{equation}
where $I_\mathrm{in}$ is the incident intensity, $\ell$ is the line-of-sight path of the beam through specimen, $I_\mathrm{out}$ is the output intensity one is able to measure. Since in practice, $\mu$ is energy dependent and typically the source is polyenergetic, the output intensity becomes
\begin{equation}
I_\mathrm{out} = \int_{\xi}I_\mathrm{in}(\xi)\exp\left(-\int_{\ell}\mu(\ell,\xi)\,d\ell\right)\,d\xi.
\end{equation}

For a finite number of photons, the measured intensity will be probabilistic with an approximate Poisson distribution \citep{Chang2014}. If we also move the attenuation, measurements and energy spectrum into a discretised setting, we can write the measurement process as
\begin{equation} \label{equ:poly-poiss}
y_i \sim \operatorname{Poisson}\left\{\sum_{j=1}^{N_\xi}b_i(\xi_j)\exp\left(-[\boldsymbol{\Phi}\boldsymbol{\mu}(\xi_j)]_i\right)+s_i \right\} \mbox{ for }  i=1,...,N_\mathrm{ray},
\end{equation}
where $N_\mathrm{ray}$ is the number of CT measurements, $N_\mathrm{\xi}$ is the number of energy bins, $\boldsymbol{b}(\xi)\in\mathbb{R}^{N_\mathrm{ray}}$ is a vector of incident intensities, $\boldsymbol{\mu}(\xi)\in\mathbb{R}^{N_\mathrm{vox}}$ is the vector of attenuation coefficients with $N_\mathrm{vox}$ the number of voxels, $\boldsymbol{\Phi}\in\mathbb{R}^{N_\mathrm{ray}\times N_\mathrm{vox}}$ is the system matrix describing the summation along the paths from source through specimen onto each detector, and $\boldsymbol{s}\in\mathbb{R}^{N_\mathrm{ray}}$ is the expectation of the scatter or other background noise reaching the detector.

\subsection{Material dependent X-ray attenuation}
The mechanism that allows various regions in a heterogeneous specimen to be differentiated is their degree of X-ray attenuation. For biological tissues irradiated with a diagnostic X-ray source, the significant phenomena contributing to the attenuation of incident radiation are photoelectric and scattering effects---consisting of Compton, Rayleigh and Thompson scatter \citep{Curry1990}. The combined attenuation strength of a given material can be quantified as
\begin{equation} \label{equ:cross_section}
\mu(\xi) = \rho N_g(\sigma_\mathrm{pe}(\xi)+\sigma_\mathrm{incoh}(\xi)+\sigma_\mathrm{coh}(\xi)),
\end{equation}
where $\xi$ is the energy of the incident radiation, $\sigma_\mathrm{pe}$, $\sigma_\mathrm{incoh}$ and $\sigma_\mathrm{coh}$ represent the interactive cross sections---quantifying the probability of interaction---of photoelectric, incoherent (Compton) and coherent (Rayleigh and Thompson) effects, $\rho$ is the mass density, and $N_g$ is the number of electrons per unit volume defined as
\begin{equation} \label{equ:elec_fraction}
N_g = N_A\sum_i\frac{\omega_iZ_i}{A_i},
\end{equation}
where $N_A$ is Avagadro's number, and $Z_i$, $A_i$, $\omega_i$ are atomic number, atomic weight and relative fraction by mass of a material's constituent elements \citep{Schneider1996}. A convenient parameter to use is the relative electron density, which is
\begin{equation} \label{equ:red}
\rho_e = \frac{\rho Ng}{\rho_\mathrm{water}N_{g,\mathrm{water}}},
\end{equation}
where $N_{g,\mathrm{water}}$ is the absolute electron density of water and $\rho_\mathrm{water}$ its mass density, having a value of almost exactly $\rho_\mathrm{water}=1\,\text{g/cm}^3$ at room temperature.

\begin{figure}[!htb]
	\centering
	\includegraphics[width=0.7\textwidth]{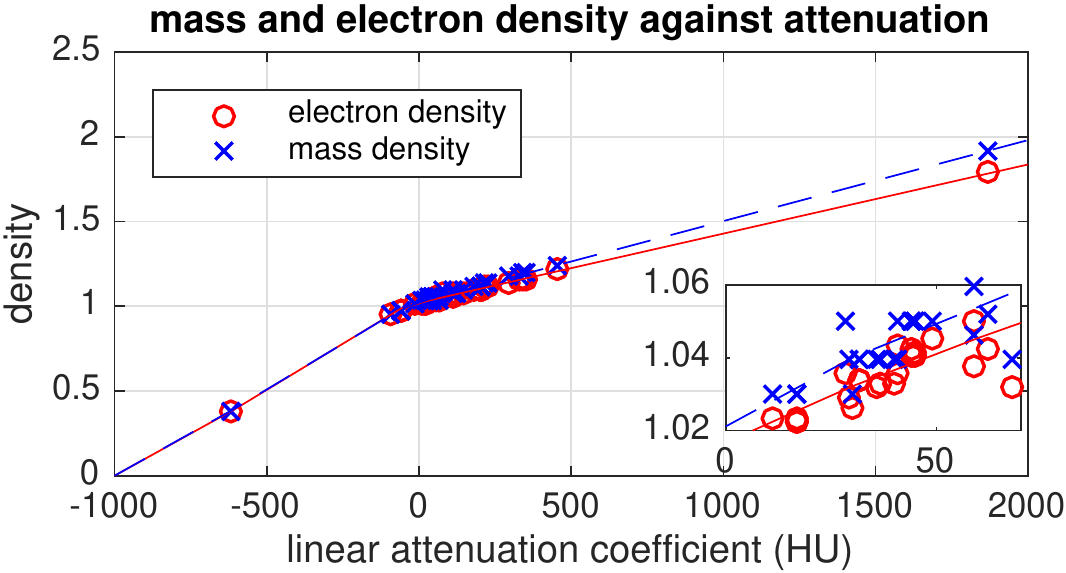}
	\caption{Plot of mass and electron density calibration curves against relative attenuation coefficient in Hounsfield Units (HU) at 60 keV, for range in biological tissues as defined in \citep{icrp2002}}
	\label{fig:hu_density}
\end{figure}

From (\ref{equ:cross_section}), (\ref{equ:elec_fraction}) and (\ref{equ:red}), one may quantify the X-ray attenuation in terms of the energy independent mass density or relative electron density---$\rho$ and $\rho_e$ respectively---given knowledge of the energy dependent interactive cross sections, which may be found from existing parameterisations in Section~\ref{sec:exist} or our proposed model in Section~\ref{sec:eden_atten}. A more common approach however, is simply to use calibration curves on the reconstructed images. Using the Hounsfield scale defined as
\begin{equation}
\text{HU}=1000\frac{\mu-\mu_\mathrm{water}}{\mu_\mathrm{water}-\mu_\mathrm{air}},
\end{equation}
examples of calibration curves are shown in Figure~\ref{fig:hu_density}.
The calibration curves in Figure~\ref{fig:hu_density} are generated from 52 human tissues defined in \citep{icrp2002} and used in \citep{icrp2009}, according to a monoenergetic X-ray source at 60 keV. As in \citep{Schneider1996}, the curves are piecewise linear, with 4 breakpoints at lung tissue, adipose tissue and around soft tissues, although these are some of these are difficult to discern in the figure. It can observed that most tissues have very similar relative electron density and mass density, which deviates more significantly with higher attenuating materials containing bone.

In order to use the calibration curve in Figure~\ref{fig:hu_density}, the attenuation should be converted to a monoenergetic equivalent, and this is often approximated prior to reconstruction as in \citep{Joseph1978}, which means performing nonlinear calibration both before and after reconstruction, to map into mass or electron density. The alternative that we consider here is to use explicit parameterisations of the attenuation.

\subsection{Existing Physical Parameterisations} \label{sec:exist}
Due to each cross-section in (\ref{equ:cross_section}) being a non-linear function of energy and material, the total attenuation of a tissue is complicated and difficult to quantify exactly. One approach is to parameterise (\ref{equ:cross_section}) as a linear combination of basis functions. For example, in \citep{Alvarez1976} a convenient choice is  
\begin{equation} \label{equ:alverez}
\mu(\xi) = \underbrace{K_1\rho_eZ_\mathrm{eff}^n\xi^{-3}}_{\text{photoelectric}} + \underbrace{K_2\rho_ef_\mathrm{KN}(\xi)}_{\text{Compton scatter}},
\end{equation}
where $f_\mathrm{KN}(\cdot)$ is the Klein--Nishina function \citep{Klein1929} describing the probability of Compton scatter, $Z_\mathrm{eff}$ is the effective atomic number for a composite material \citep{Weber1969}, and $K_1$, $K_2$ and $n$ are scalar parameters to fit the model to data. It should be noted that for unbound electrons, one would have $\sigma_\mathrm{incoh}=f_\mathrm{KN}(\cdot)$. Additionally, the coherent scattering events are not explicitly modelled, though their contribution is low at the energies of interest \citep{Jackson1981}.

\begin{figure}[!htb]
	\centering
	\begin{subfigure}[b]{0.49\textwidth}
		\includegraphics[trim=0.5cm 0cm 1cm 0cm,clip=true,width=\textwidth]{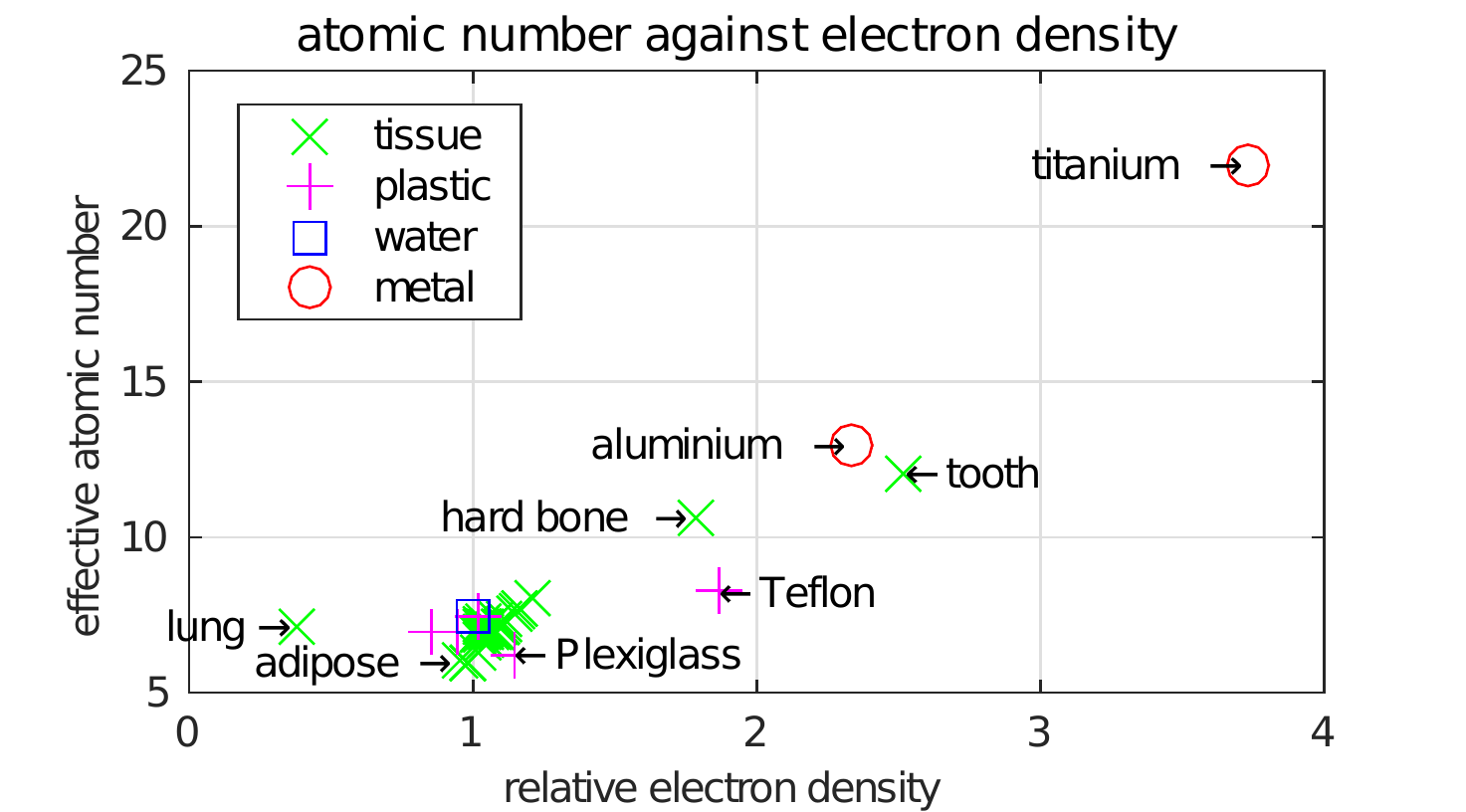}
		\caption{}
		\label{subfig:pe_z}
	\end{subfigure}
	\begin{subfigure}[b]{0.49\textwidth}
		\includegraphics[trim=0.5cm 0cm 1cm 0cm,clip=true,width=\textwidth]{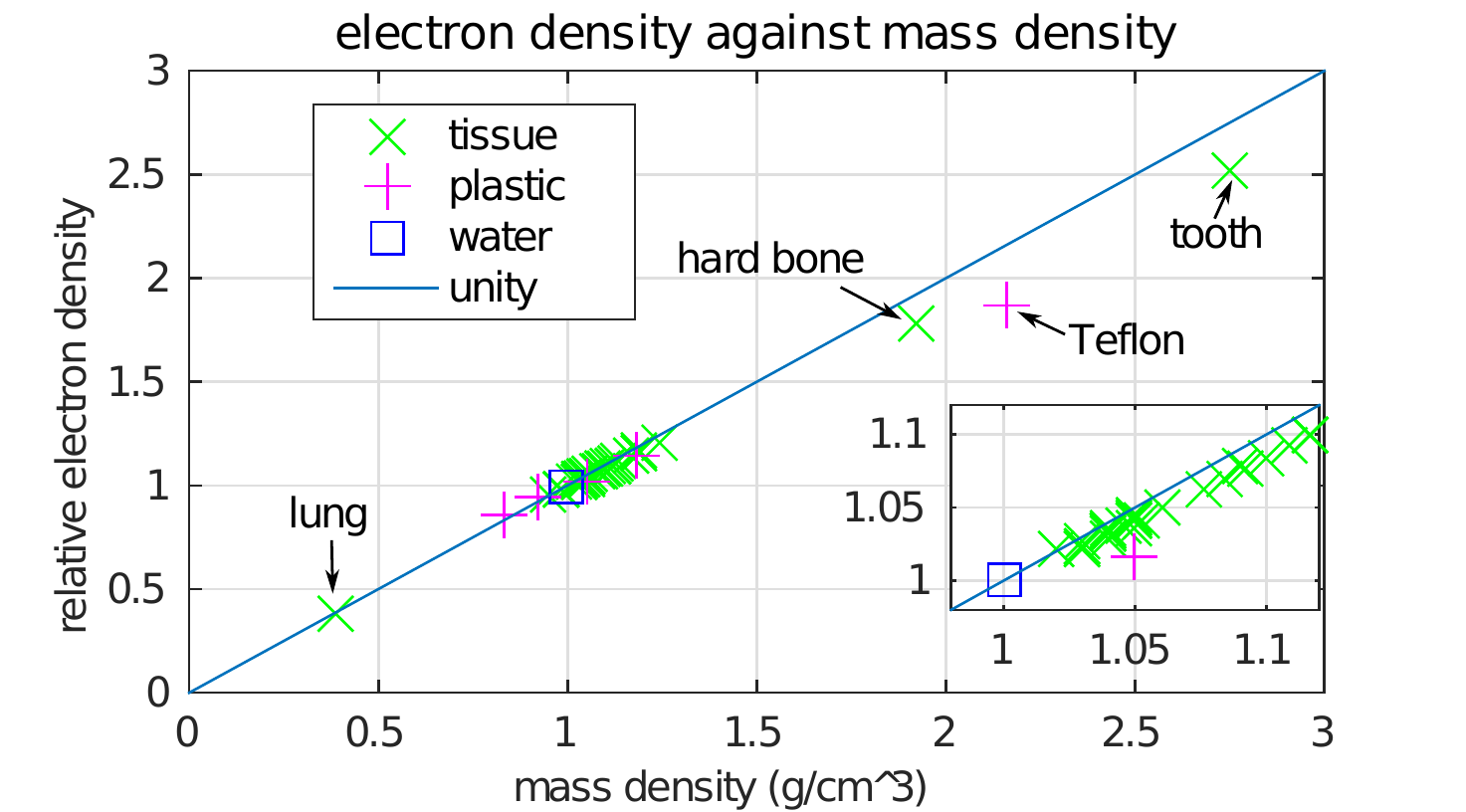}
		\caption{}
		\label{subfig:den_eden}
	\end{subfigure}
	\caption{(a) Plot of effective atomic number against relative electron density for a materials of: tissues from \citep{icrp2002}, plastics, water and metals; (b) relation between relative electron density and mass density for tissue, plastics and water, with unity line shown for illustration}
	\label{fig:pe_z}
\end{figure}

According to (\ref{equ:alverez}), the energy dependent attenuation of any material may be parameterised by its effective atomic number and electron density. Several examples of these parameters for a range of differing material types are plotted in Figure~\ref{subfig:pe_z}.
Although there is clearly a strong correlation between the relative electron density $\rho_e$ and effective atomic number $Z_\mathrm{eff}$, in order to unambiguously model all classes of materials, at least two spectral measurements must be taken to separate the contribution from the two terms in (\ref{equ:alverez})---since there is no one-to-one mapping from $\rho_e$ to $Z_\mathrm{eff}$. This is the basis of a DECT technique, where from the model in (\ref{equ:alverez}) and measurements from two distinct X-ray spectra, the attenuation can be decomposed into photoelectric and Compton, from which $\rho_e$ could be unambiguously calculated \citep{Alvarez1976}.

In \citep{DeMan2001a}, the authors use a piecewise-linear fit from a monoenergetic equivalent attenuation to both of the energy independent factors in (\ref{equ:alverez}): $K_1\rho_eZ_\mathrm{eff}^n$ for photoelectric attenuation; and $K_2\rho_e$ for Compton scatter. This fitting allows polyenergetic reconstruction from a single source. Due to the degradation of the energy dependent modelling in (\ref{equ:alverez}) at higher effective atomic number or higher energy however \citep{Jackson1981}, this will not be quantitatively consistent in $\rho_e$ between hard and soft materials; the model is also shown not to be consistent between synthetic and biological materials \citep{DeMan2001a}. 

Another idea is to use physical materials as basis functions. For example, for biological specimens, water and bone may be considered \citep{Joseph1978}. The attenuation is then
\begin{equation} \label{equ:water_bone}
\mu(\xi) = a_1\mu_\mathrm{water}(\xi)+a_2\mu_\mathrm{bone}(\xi) = \rho\left(a_1m_\mathrm{water}(\xi)+a_2m_\mathrm{bone}(\xi)\right),
\end{equation}
where $\rho$ is the mass density, $m_\mathrm{water}(\xi)$ and $m_\mathrm{bone}(\xi)$ are the energy dependent mass attenuation coefficients, and $a_1$ and $a_2$ can be binary \citep{Elbakri2002} or water--bone fractions \citep{Elbakri2003}.

One may calculate $\rho$ and map into $\rho_e$ through a non-linear calibration if desired. The relation between mass and electron density is shown in Figure~\ref{subfig:den_eden}.
Although higher density materials deviate from that of water, as long as one can generate an estimate of the mass density, then the trend shown in Figure~\ref{subfig:den_eden} can be approximated as piecewise-linear and conversion to electron density is possible, which is similar to the HU against density plot shown in Figure~\ref{fig:hu_density}. A possible weakness of this model is the inaccuracies that will occur when tissues have a dissimilar mass-attenuation profile to both water or hard bone, such as adipose tissue \citep{Schneider1996}.

\section{Methodology} \label{sec:method}
\subsection{Polyquant Attenuation Model} \label{sec:eden_atten}
\begin{figure}[!htb]
	\centering
	\begin{subfigure}[b]{0.6\textwidth}
		\includegraphics[width=\textwidth]{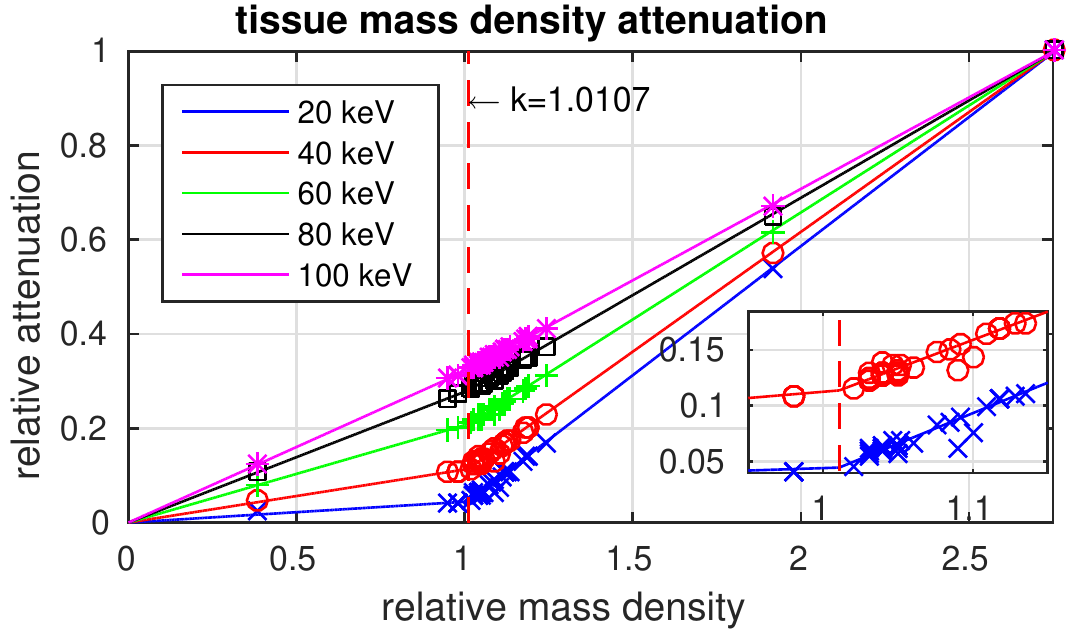}
	\end{subfigure}
	\hfil
	\begin{subfigure}[b]{0.6\textwidth}
		\includegraphics[width=\textwidth]{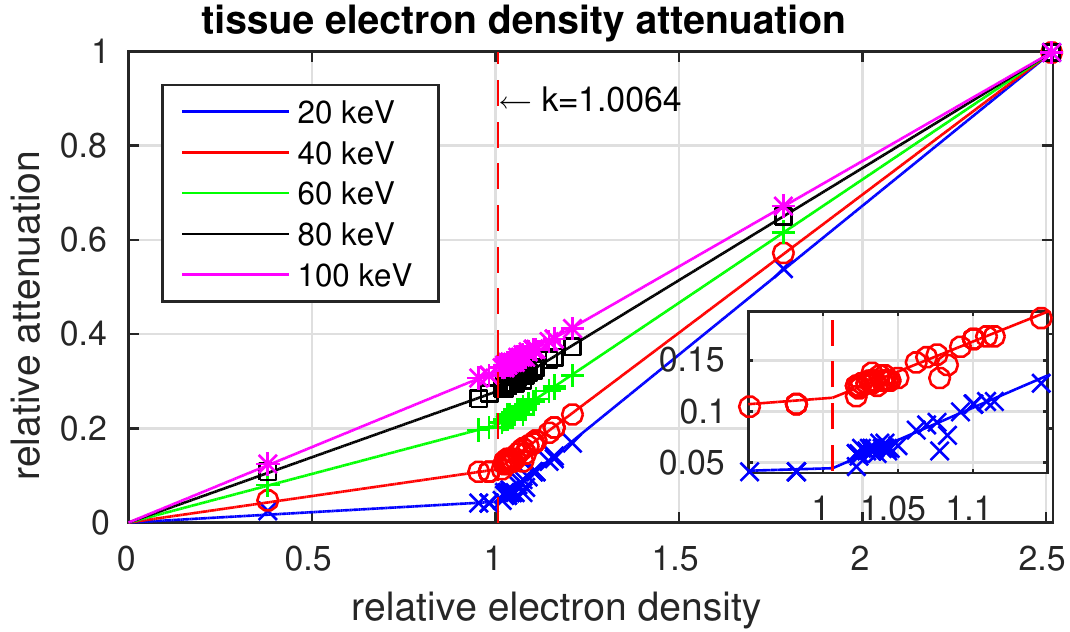}
	\end{subfigure}
	\caption{Relation between energy dependent attenuation and relative electron and mass density for biological tissues from \citep{icrp2002}. The attenuation in each case is normalised to that of tooth, to allow visualisation on a single graph}
	\label{fig:pe_atten}
\end{figure}

We propose to take a data centric rather than a physical approach to parameterising the X-ray attenuation. To motivate our choice for this, we have plotted the relative attenuation coefficient against electron and mass density for 52 biological tissues in \citep{icrp2002} at a number of energies in Figure~\ref{fig:pe_atten}. We have normalised the attenuation to the maximum for an energy---the tooth in each case---simply to allow visualisation on a single graph.

It appears from Figure~\ref{fig:pe_atten} that the chosen biological tissues approximately follow an energy dependent two part linear fit in both electron and mass density. We have superimposed such a fit onto each plot in Figure~\ref{fig:pe_atten}, where the transition point $k$ is fixed to be the same for each energy and set to minimise the norm of the residual through all the data. Imposing a constant $k$ throughout will motivate computational efficiency outlined in the following section. One interpretation of this parameterisation is taking the inverse of HU to density shown in Figure~\ref{fig:hu_density} and incorporating a dependency on energy.

Generalising the piecewise linear fit motivated from the data in Figure~\ref{subfig:den_eden} may be written as
\begin{equation} \label{equ:gen_fit}
\hat{\boldsymbol{\mu}}(\boldsymbol{x},\xi) = \sum_{i=1}^{N_f}\boldsymbol{f}_i\odot(\alpha_i(\xi)\boldsymbol{x}+\beta_i(\xi)), 
\end{equation}
where $N_f$ is the number of linear fits, $f_i$ is a class indicator function for materials belonging to the $i^{th}$ class, and $\boldsymbol{x}\in\mathcal{R}^{N_\mathrm{vox}}$ is the physical parameter one wishes to fit to---electron density or mass density for example. The class indicator functions are defined as
\begin{equation}
f_i = 
\begin{cases} 
1 & \text{if } k_{i-1} \leq x < k_i\\
0       & \text{otherwise}
\end{cases}
\mbox{ for }  k=1,\dots,N_f,
\end{equation}
where $k_i$ is the `knee' between the $i^{th}$ and $(i+1)^{th}$ linear fits, with $k_0=0$. To ensure connected fits, we may also enforce the constraints
\begin{equation} \label{equ:cont}
\alpha_{i}k_{i}+\beta_i=\alpha_{i+1}k_i+\beta_{i+1}.
\end{equation}
We also have $k_{N_f} = 0$ and $\beta_1 = 0$, which ensures that a vacuum has no attenuation. 

Following on from the connection to the decomposed inverse of calibration in Figure~\ref{fig:hu_density}, this general model would indeed allow several breakpoints around different tissue classes as in \citep{Schneider1996}, but here we focus on two or three for computational simplicity.

\subsubsection{Special Cases and Connections to Existing Models} \label{sec:model_exist}
We now give several interesting special cases and connections to existing models:
\begin{itemize}
\item
An example of (\ref{equ:gen_fit}) may be used for parameterising the attenuation of biological tissues as a function of relative electron density, which was the singular case studied in \citep{Mason2017}, and allows the fitting in Figure~\ref{fig:pe_atten}. For this we have $N_f=2$, $x=\rho_e$ and $k=1.0064$, which may be written as
\begin{equation} \label{equ:logit-fit}
\hat{\boldsymbol{\mu}}(\boldsymbol{\rho_e},\xi) = \boldsymbol{f}_1(\boldsymbol{\rho_e})\odot \alpha_1(\xi)\boldsymbol{\rho_e}+\boldsymbol{f}_2(\boldsymbol{\rho_e})\odot [\alpha_2(\xi)\boldsymbol{\rho_e}+\beta_2(\xi)].
\end{equation}
This is the instance we study, along with an equivalent fit for mass density. 

\item
The model above may also include highly dense materials, such as titanium metallic implants with a mass density of $\rho=4.5$ g/cm$^3$. To include this, a second knee point may be inserted between bone and metal, and one would have $N_f=3$. We also study this in our numerical experiment with hip implants.

\item
Another interesting special case with (\ref{equ:gen_fit}) is when all $b_i$ are set to 0, $x=\rho$, $N_f=2$, $\alpha_1(\xi)$ and $\alpha_2(\xi)$ are mass attenuation coefficients for water and bone, and the continuity constraints are ignored. This yields (\ref{equ:water_bone}), with a density dependent threshold, and is equivalent to the formulation in \citep{Elbakri2003} without smoothing. The consequence of ignoring the continuity in (\ref{equ:cont}) will be a step at the `knee' points.

\item
In a similar manner to the water--bone model utilised in \citep{Elbakri2003}, the IMPACT model in \citep{DeMan2001a} may also be expressed as a special case of our general fitting model in (\ref{equ:gen_fit}). We note that adopting the same notation, the model may be written as
\begin{equation} \label{equ:de_man}
\hat{\boldsymbol{\mu}}(\boldsymbol{x},\xi) = \sum_{i=1}^{N_f}\boldsymbol{f}_i\odot\left(\left[\frac{\alpha_i}{\xi^3}+\gamma_if_\mathrm{KN}(\xi)\right]\boldsymbol{x}+\frac{\beta_i}{\xi^3}+\delta_if_\mathrm{KN}(\xi)\right),
\end{equation}
where $\alpha_i$, $\beta_i$, $\gamma_i$ and $\delta_i$ are the coefficients or photoelectric and Compton linear fits, and $\boldsymbol{x}$ is a monoenergetic equivalent attenuation. A key difference between (\ref{equ:de_man}) and (\ref{equ:gen_fit}) is the energy dependence of coefficients and number of parameters. Whilst our model has $2N_fN_\xi-1$ free parameters, (\ref{equ:de_man}) has $4N_f$, which is likely to be less, thereby offering increased model flexibility but, as we will see in the next section, at no additional computational cost.

\item
We could also use (\ref{equ:gen_fit}) to predict the reconstruction of a monoenergetic equivalent image---in DECT this is known as `quasi monoenergetic' reconstruction \citep{Johnson2012}. This is possible since the imposition of a constant `knee' position across energies allows any monoenergetic attenuation to be predicted using an equivalent piecewise linear function. The relation to the IMPACT model in this case  is that the energy dependence is fitted from the data instead of implied from the photoelectric--Compton model in (\ref{equ:alverez}).

\item
Although not evaluated in this article, due to the same shape in HU to proton stopping power in \citep{Schneider1996} as relative electron density, we suggest our model should be very applicable for proton interaction modelling also.
\end{itemize}

\subsection{Direct Quantitative Density Reconstruction} \label{sec:quant_recon}
The Polyquant model introduced in Section~\ref{sec:eden_atten} describes the forward mapping from physical quantity to attenuation. We now show how this may be combined with the statistical CT measurement model in (\ref{equ:poly-poiss}), to allow direct statistical inference of mass or electron density. 

Combining (\ref{equ:gen_fit}) with (\ref{equ:poly-poiss}) results in the relation
\begin{equation}
\sum_{j=1}^{N_\xi}b_i(\xi_j)\exp\left(-[\boldsymbol{\Phi}\boldsymbol{\mu}(\xi_j)]_i\right) = \sum_{j=1}^{N_\xi}b_i(\xi_j)\exp\left(-[\boldsymbol{\Phi}\boldsymbol{\hat{\mu}}(\boldsymbol{x},\xi_j)]_i\right) \mbox{ for }  i=1,...,N_\mathrm{ray}.
\end{equation}
If we introduce a function $\psi(\cdot,\cdot)$ to simplify notation as
\begin{equation}
\psi_i(\boldsymbol{x},\xi) \equiv b_i(\xi)\exp\left(-[\boldsymbol{\Phi}\boldsymbol{\hat{\mu}}(\boldsymbol{x},\xi)]_i\right) \mbox{ for }  i=1,...,N_\mathrm{ray},
\end{equation}
we can write the negative log-likelihood (NLL) for the Poisson model as
\begin{equation} \label{equ:nll}
\mathrm{NLL}(\boldsymbol{x};\boldsymbol{y}) =
\sum_{i=1}^{N_\mathrm{ray}}\sum_{j=1}^{N_\xi}\psi_i(\boldsymbol{x},\xi_j)+s_i - y_i\log\left(\sum_{j=1}^{N_\xi}\psi_i(\boldsymbol{x},\xi_j)+s_i\right),
\end{equation}
where we note that this function is non-convex as with similar CT NLL functions in \citep{Erdogan1999,Chang2014}.

Reconstruction of the quantitative density map can be performed by finding an $\boldsymbol{x}$ that minimises (\ref{equ:nll}). We will look at gradient descent methods, for which we require an expression for the derivative of NLL. If we simplify notation with the following
\begin{equation}
\boldsymbol{d}(\boldsymbol{x}) = \boldsymbol{y}\oslash\left(\sum_{j=1}^{N_\xi}\boldsymbol{\psi}(\boldsymbol{x},\xi_j)+\boldsymbol{s}\right) - \mathbf{1},
\end{equation}
where $\oslash$ represents component-wise division. An expression for the derivative is then
\begin{equation} \label{equ:deriv}
\frac{\partial\mathrm{NLL}(\boldsymbol{x};\boldsymbol{y})}{\partial\boldsymbol{x}} \approx
\sum_{i=1}^{N_f}\boldsymbol{f}_i(\boldsymbol{x})\odot\boldsymbol{\Phi}^T\left[\sum_{j=1}^{N_\xi}\alpha_i(\xi_j)\boldsymbol{\psi}(\boldsymbol{x},\xi_j)\odot\boldsymbol{d(\boldsymbol{x})}\right],
\end{equation}
where $\boldsymbol{\Phi}^T$ represents a transpose of the system matrix or `back-projection', and $\odot$ is component-wise multiplication. We have shown this derivative as an approximation ``$\approx$'' in testament to the fact that there are discontinuities in the gradient at the `knee' positions, where the gradient is not defined. Although in \citep{Mason2017}, we invoked the logistic function to mitigate this effect, we have found better empirical performance by instead using (\ref{equ:deriv}) directly.

Before we proceed, we note that (\ref{equ:deriv}) only has a single backprojection operation $\boldsymbol{\Phi}^T$ per linear fit, so the number of these operations is independent of the number of energies $N_\xi$. This is only possible due to constant `knee' positions $k$ for each energy. Calculating $\boldsymbol{\Phi}\boldsymbol{\hat{\mu}}(\boldsymbol{x},\xi)$ is also independent on $N_\xi$, and can be evaluated with $2N_f-1$ applications of $\boldsymbol{\Phi}$, where one fewer is a consequence of the assumption that $\beta_1=0$. For example, applying the method for electron density reconstruction of tissues using (\ref{equ:logit-fit}), one would expect a computation cost of three forward- and two backprojections.

Although one may obtain a quantitative density reconstruction through maximum likelihood estimation, by iteratively minimising (\ref{equ:nll}) through gradient descent with (\ref{equ:deriv}), incorporation of prior regularisation will typically improve statistical performance, especially as the noise increases or when few measurements are taken. Incorporating regularisation now gives the penalised log-likelihood or maximum a posteriori estimate as
\begin{equation} \label{equ:obj}
\hat{\boldsymbol{x}} = \argmin_{\boldsymbol{x}\in\mathcal{C}}\mathrm{NLL}(\boldsymbol{x};\boldsymbol{y}) + \lambda R(\boldsymbol{x}),
\end{equation}
where $R(\cdot)$ is some regularisation function, and $\mathcal{C}$ is a set of box constraints on $\boldsymbol{x}$ so that $0\leq x_i\leq \zeta \mbox{ for }  i=1,...,N_\mathrm{vox.}$, where $\zeta$ is the maximum allowable density value, and the constraint set ensures non-negative density values.

The choice of $R(\cdot)$ in (\ref{equ:obj}) will vary on the imaging application, but some possibilities are generalized Gaussian Markov random field (GGMRF) \citep{Bouman1993}, total variation (TV) \citep{Rudin1992} or wavelet sparsity \citep{Daubechies2008}. Without loss of generality, we opt in our experimental section for TV, since it promotes piecewise flat images, which we expect from homogeneous slabs of tissue.

\subsubsection{Algorithm Design} \label{sec:algorithm}
Although there are many approaches for evaluating (\ref{equ:obj}), we will detail here the algorithm we have used in our experiments, which we give in Algorithm~\ref{alg1}.
\begin{algorithm}                      
	\caption{Prox-Polyquant}
	\label{alg1}
	\begin{algorithmic}                    
		\STATE $\gamma \leftarrow [0,1]$
		\STATE $\boldsymbol{x}^0 \leftarrow \mathbf{1}$
		\STATE $\delta \leftarrow 2(1-\gamma)/L_0$
		\STATE $\boldsymbol{x}^1 \leftarrow \boldsymbol{x}^0$
		\FOR{$k = 1,2,\ldots$}
		\STATE $\boldsymbol{x}^{k+1} \leftarrow \mathbf{prox}_{\delta R}\left(\boldsymbol{x}^{k}-\delta\frac{\partial\mathrm{NLL}(\boldsymbol{x}^{k};\boldsymbol{y})}{\partial\boldsymbol{x}^{k}}+\gamma(\boldsymbol{x}^{k}-\boldsymbol{x}^{k-1})\right)$
		\ENDFOR
	\end{algorithmic}
\end{algorithm}

We note that Algorithm~\ref{alg1} is an instance of the iPiano \citep{Ochs2014}, although we have made some slight changes, where we chose this method due to its analysis with non-convex objective functions as we have in (\ref{equ:nll}). The parameter $\delta$ is the step size, which scales with a factor of $L_0$, which we define as
\begin{equation}
L_0 = \left\|\boldsymbol{\Phi}^T\left[\left(\sum_{j=1}^{N_\xi}\alpha_1(\xi_j)\boldsymbol{b}(\xi_j)\right)\odot\boldsymbol{\Phi1}\right]\right\|_\infty,
\end{equation}
where $\|\cdot\|_\infty$ is the maximum norm of the vector. $L_0$ represents the maximum in the diagonal of the Hessian of (\ref{equ:deriv}) at the point $\boldsymbol{0}$, and may be considered as a crude estimate of the global Lipschitz constant, $L$. We note that this step size is likely to be very conservative in practice. The parameter $\gamma$ sets the `inertia' of the method, where we used $\gamma=0.8$ as resulted in the fastest performance in \citep{Ochs2014}. Finally, $\mathbf{prox}_{\delta R}$ is the proximity operator defined as
\begin{equation} \label{equ:prox}
\mathbf{prox}_{\delta R}(\boldsymbol{z}) = \argmin_{\boldsymbol{\rho}\in\mathcal{C}} \frac{1}{2} \|\boldsymbol{z}-\boldsymbol{\rho}\|_2^2 + \delta\lambda R(\boldsymbol{\rho}).
\end{equation}
In our experimental section, we use the TV \citep{Rudin1992} as a regularisation function $R$, in which case (\ref{equ:prox}) may be evaluated as in \citep{Beck2009a}.

\begin{figure}[!htb]
	\centering
	\includegraphics[width=0.7\textwidth]{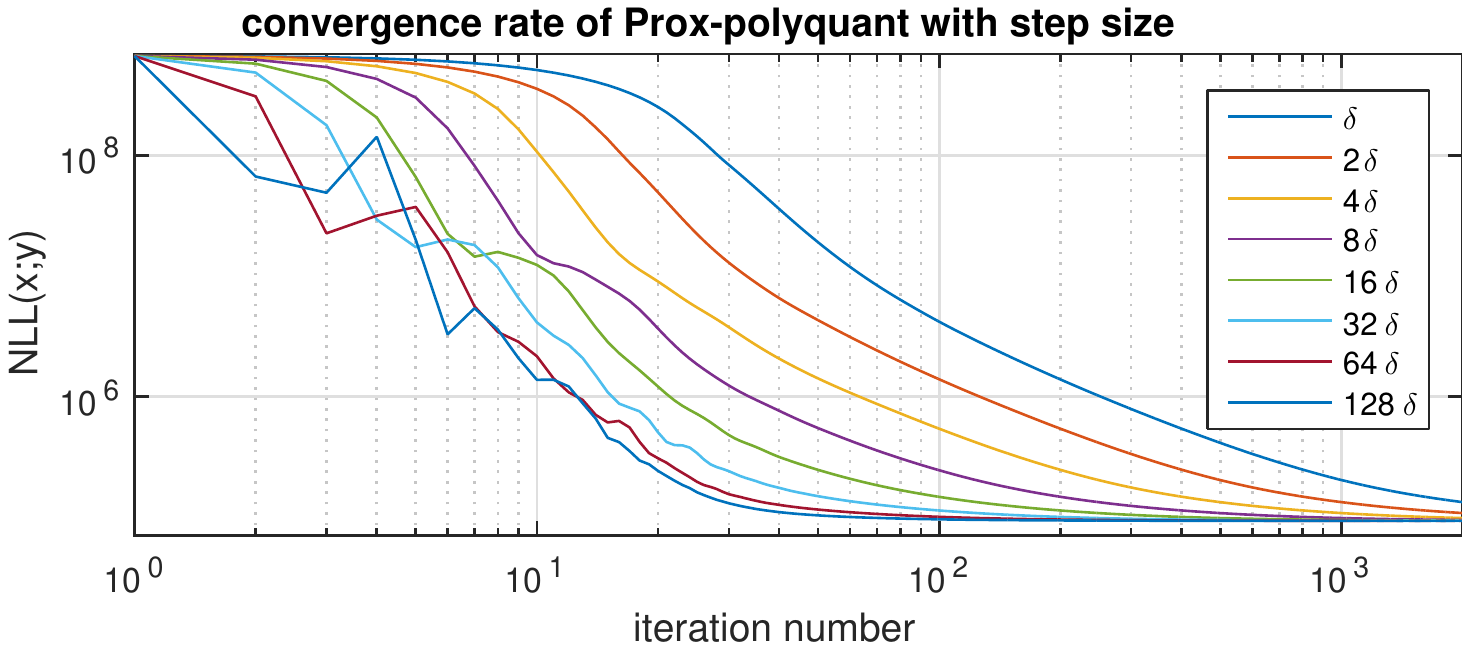}
	\caption{Convergence analysis for various step size multiplication factors without regularisation applied to the \textit{chest data} detailed in Section~\ref{sec:rec_test}. Each NLL here has a constant value added to it to aid visualisation}
	\label{fig:converge}
\end{figure}

To illustrate the convergence properties of our method, and investigate its robustness to more aggressive step sizes, we applied Algorithm~\ref{alg1} to the \textit{chest data} detailed in Section~\ref{sec:rec_test}, and plotted the evolution of the NLL through iterates in Figure~\ref{fig:converge}, for different step size multiplication factors.
It follows that using $\delta$ is indeed very pessimistic, and one observes convergence in this case even with $128\delta$, where convergence is achieved around 100 iterations. For factors larger than 128, we have found the method fails to converge, and it can be seen than the objective does dot decrease monotonically for more aggressive step sizes. This analysis suggests that the local Lipschitz constant of the gradient through iterations is normally significantly less than the global maximum, or indeed the approximation with $L_0$. Adaptive step size schemes could exploit this fact, such as the backtracking line-search in \citep{Ochs2014}, but one should be aware that evaluating the cost function in (\ref{equ:obj}) will be more than half the cost of calculating a new gradient, so each backtrack evaluation will increase the iteration cost significantly.

\section{Experimentation} \label{sec:experiment}
For our experiments, we investigate the model accuracy in Section~\ref{sec:model_test} followed by a numerical fan-beam CT quantitative reconstruction test in Section~\ref{sec:rec_test}, and finally validation with real data in Section~\ref{sec:ccbt_recon}.

\subsection{Polyquant Model Evaluation} \label{sec:model_test}
To investigate the accuracy of our proposed attenuation model, and its comparison to the other parameterisations given in Section~\ref{sec:exist}, we calculated the predicted linear attenuation coefficient for a number of representative tissues from ICRP 89 \citep{icrp2002}: adipose tissue, muscle tissue, spongy tissue (upper femoral spongiosa), and hard bone.

\begin{figure}[!htb]
	\centering
	\begin{subfigure}[b]{0.45\textwidth}
		\includegraphics[width=\textwidth]{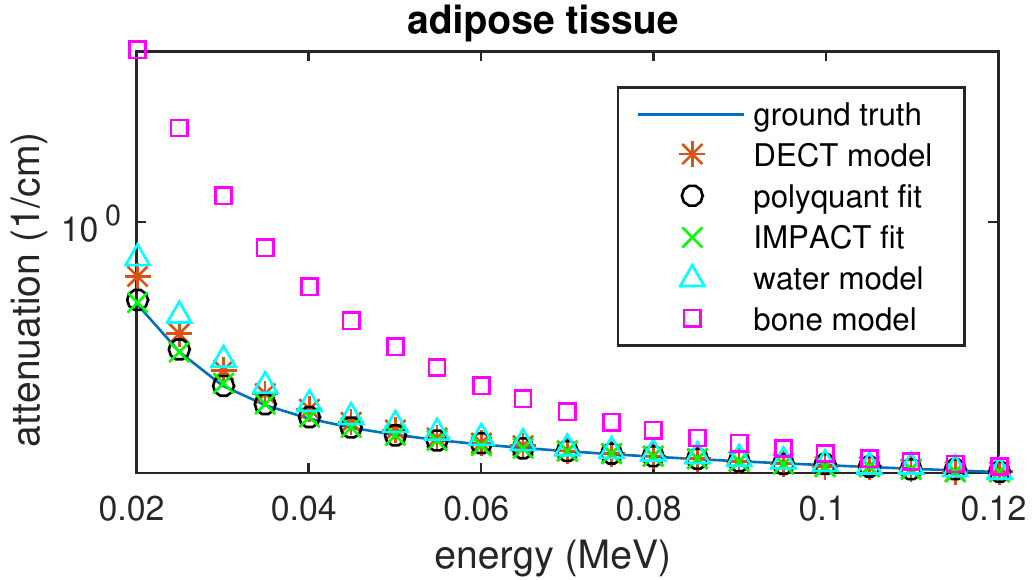}
	\end{subfigure}
	\hfil
	\begin{subfigure}[b]{0.45\textwidth}
		\includegraphics[width=\textwidth]{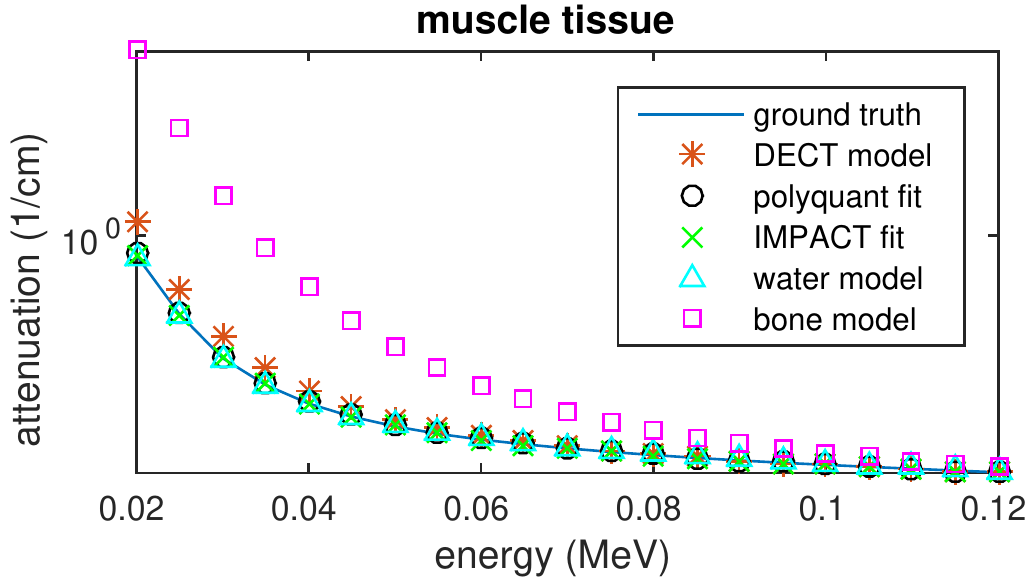}
	\end{subfigure}
	\vfill
	\begin{subfigure}[b]{0.45\textwidth}
		\includegraphics[width=\textwidth]{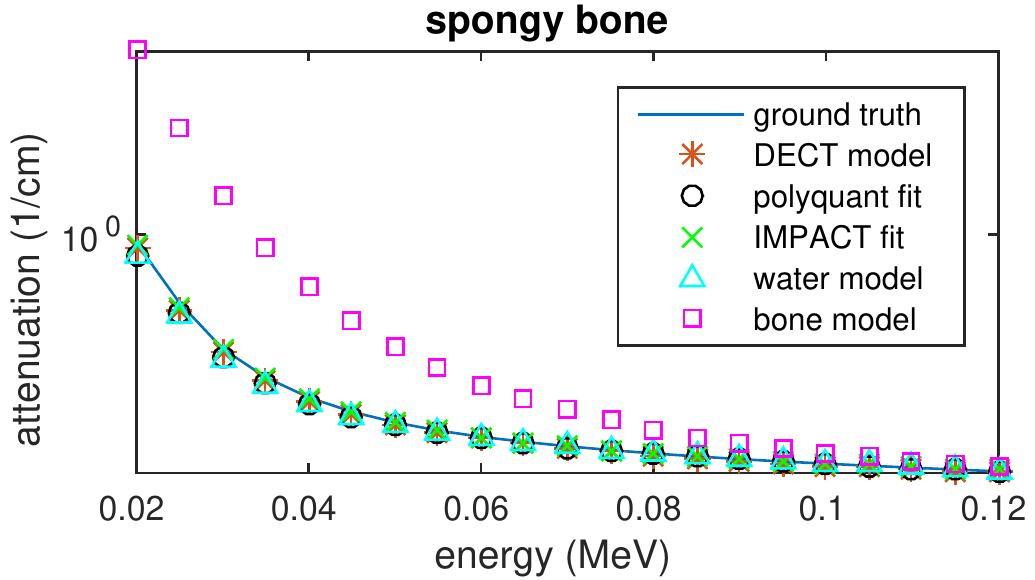}
	\end{subfigure}
	\hfil
	\begin{subfigure}[b]{0.45\textwidth}
		\includegraphics[width=\textwidth]{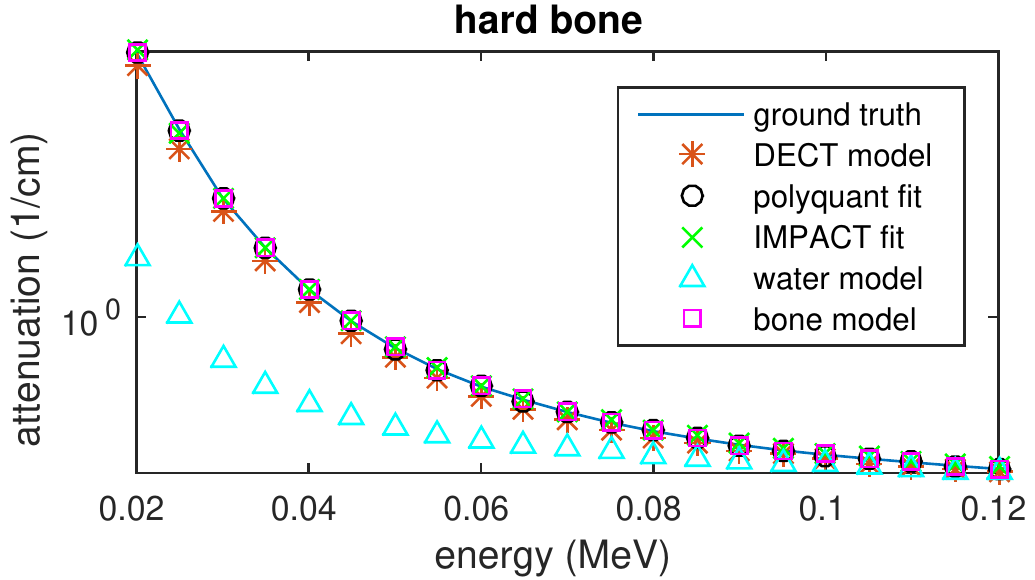}
	\end{subfigure}
	\caption{Model fitting results for tissue materials: adipose tissue, muscle tissue, spongy bone (upper femoral spongiosa), and hard (cortical) bone.}
	\label{fig:tissue_fit}
\end{figure}

For the DECT model in (\ref{equ:alverez}), we optimised the three scalar parameters $K_1$, $K_2$ and $n$ for best fit to all materials in the ICRP 89---shown in Figure~\ref{fig:pe_atten}. We also fitted our piecewise linear Polyquant model in Section~\ref{sec:eden_atten} for both relative electron density and monoenergetic attenuation at 60 keV to the same data. Similarly, we evaluated the accuracy of the IMPACT parameterisation in \citep{DeMan2001a} to monoenergetic attenuation, again using the same fitting data from ICRP 89. These are plotted along with water and bone attenuation models in Figure~\ref{fig:tissue_fit} and the residual norms are tabulated in Table~\ref{tab:tissue_res}.
\begin{table}[!htb]
	\caption{Residual norm for various models fitted to various tissue materials.}
	\label{tab:tissue_res}
	\centering
	$
	\begin{array}{c|c|c|c|c|c}
	\text{Scheme} & \text{adipose tissue} & \text{muscle tissue} & \text{spongy bone} & \text{hard bone}\\
	\hline
	\text{DECT} & 0.15 & 0.28 & 0.047 & 0.97\\
	\text{Water} & 0.27 & 0.015 & 0.12 & 6.9\\
	\text{Bone} & 3.7 & 3.8 & 3.7 & \mathbf{0.0}\\
	\text{IMPACT} & 0.029 & 0.011 & 0.038 & 0.15\\
	\text{Polyquant-tissue-}\rho_e & 0.011 & 0.018 & 0.11 & 0.031\\
	\text{Polyquant-tissue-}\mu_\mathrm{mono} & \mathbf{0.010} & \mathbf{0.0016} & \mathbf{0.037} & 0.0080\\
	\hline
	\end{array}
	$
\end{table}

An observation that can be made from the plots in Figure~\ref{fig:tissue_fit} and results in Table~\ref{tab:tissue_res} are that both IMPACT and our Polyquant models are very accurate over all materials. Fitting from monoenergetic attenuation as in IMPACT or `Polyquant-tissue-$\mu_\mathrm{mono}$' is understandably more accurate than $\rho_e$, due to its closer similarity to other monoenergetic attenuations, and one will inevitably lose this accuracy if one later calibrates to $\rho_e$ or $\rho$ using the trend in Figure~\ref{fig:hu_density}. We note that using our energy dependent fitting strategy in the case of monoenergetic attenuation consistently outperforms IMPACT, and this difference is an order of magnitude for the case of hard bone.

Another feature of the numerical results in Table~\ref{tab:tissue_res} is that explicitly using the photoelectric--Compton relation in (\ref{equ:alverez}) as in DECT is less accurate than fitting it to biological materials as in IMPACT. This highlights that the physical photoelectric--Compton model is not very accurate over a wide range of material types, but also that one should not expect the fitted parameters from IMPACT to necessarily be physically meaningful in terms of $\rho_e$ and $Z_\mathrm{eff}$.

Finally, the water model is reasonably accurate for the soft tissues and spongy bone and not hard bone, and the bone model conversely so, which suggests the component-wise model in (\ref{equ:water_bone}) is sensible. However, even if one selected the best cases from bone or water, as is the essence of \citep{Joseph1978,Elbakri2002,Elbakri2003}, then the error would still be considerably higher than either IMPACT or our proposed approach.

To demonstrate that these fitted parameters are \emph{not} universal, we also ran the same models with the same fitted parameters on two synthetic plastic materials: Plexiglass\textsuperscript{\textregistered}/acrylic (polymethyl methacrylate) and Teflon\textsuperscript{\textregistered} (polytetrafluoroethylene). These are interesting materials since they have similar electron densities to muscle and bone respectively but significantly different attenuation. In order to extend these models to include metallic implants, we also looked at how well they may model the attenuation of solid Titanium, where both IMPACT and Polyquant included a second `knee' point to incorporate its attenuation also.
\begin{figure}[!htb]
	\centering
	\begin{subfigure}[b]{0.45\textwidth}
		\includegraphics[width=\textwidth]{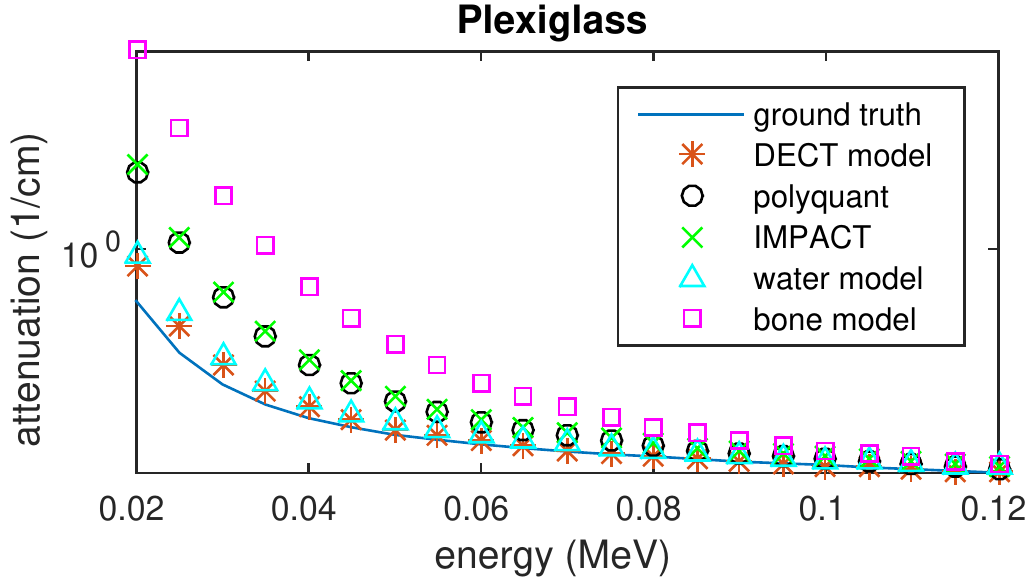}
	\end{subfigure}
	\hfil
	\begin{subfigure}[b]{0.45\textwidth}
		\includegraphics[width=\textwidth]{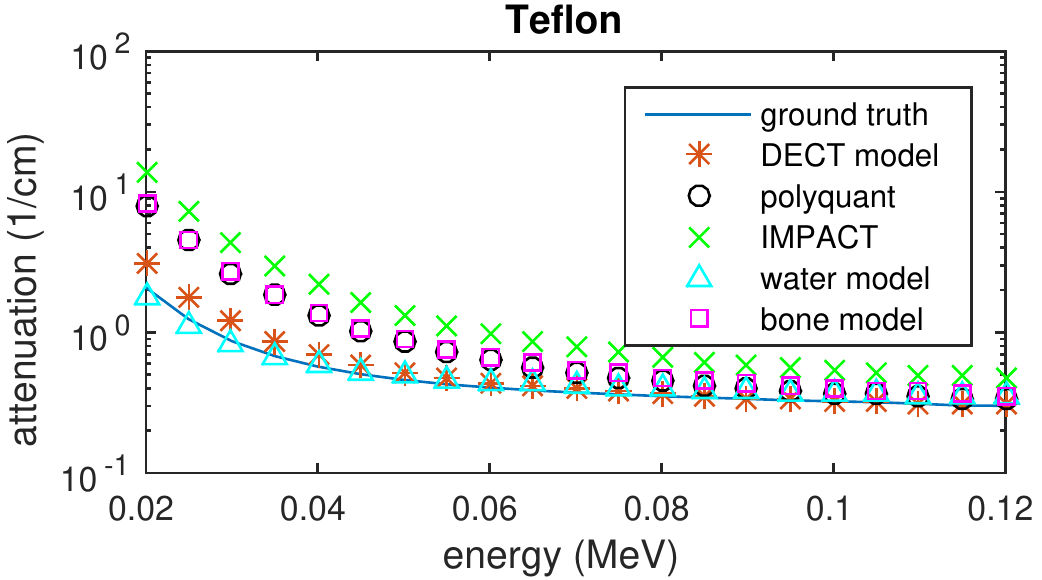}
	\end{subfigure}
	\caption{Model fitting results for synthetic plastic materials: Plexiglass\textsuperscript{\textregistered} (acrylic) and Teflon\textsuperscript{\textregistered}.}
	\label{fig:plastic_fit}
\end{figure}
\begin{table}[!htb]
	\caption{Residual norm for various models fitted to Plexiglass\textsuperscript{\textregistered} (acrylic) and Teflon\textsuperscript{\textregistered}.}
	\label{tab:platic_res}
	\centering
	$
	\begin{array}{c|c|c|c|c}
	\text{Scheme} & \text{Plexiglass} & \text{Teflon} & \text{Titanium}\\
	\hline
	\text{DECT} & 0.24 & 1.3 & 79\\
	\text{Water} & 0.35 & 0.40 & 82\\
	\text{Bone} & 4.6 & 7.5 & 66\\
	\text{IMPACT} & 1.5 & 14 & 1.9\\
	\text{Polyquant-tissue-}\rho_e & 1.3 & 7.2 & 0.032\\
	\text{Polyquant-tissue-}\mu_\mathrm{mono} & 1.5 & 14 & \mathbf{0.0076}\\
	\hline
	\text{Polyquant-plastic-}\rho_e & \mathbf{0.020} & \mathbf{0.0028} & 84\\
	\hline
	\end{array}
	$
\end{table}

From the plots in Figure~\ref{fig:plastic_fit} and results in Table~\ref{tab:tissue_res}, it is apparent that our proposed model is not universal across material types. Indeed, a water model is the best performing upon the specific plastic case, and the DECT is the best universal model, though this is also to be expected given the data in Figure~\ref{fig:pe_z} since it uses a two-dimensional parameterisation.

For the metallic implants, we note how the DECT, water and bone models are very inaccurate. IMPACT also shows significant errors, despite fitting the photoelectric--Compton model directly to Titanium, which provides further evidence that the energy dependence model is not universal. On the other hand, by imposing no physical model, our Polyquant approaches are able to capture the attenuating profile of the implant material very closely.

The implications from the model experiment is that although our model is able to fit very closely to both tissue materials and metallic implants, once these parameters are fitted, they are inaccurate in synthetic plastic materials. An explanation for this mismatch is due to the fact that the correlation between electron density and attenuation coefficient is significantly different with tissues and synthetic materials \citep{Schneider1996}. With this, although parameterising the mass attenuation coefficient with water and bone gives more consistent predictions across material types, if one wishes to map into electron density for radiation therapy applications, then one must also take care to fit to the appropriate class of materials.
\begin{figure}[!htb]
	\centering
	\begin{subfigure}[b]{0.49\textwidth}
		\includegraphics[width=\textwidth]{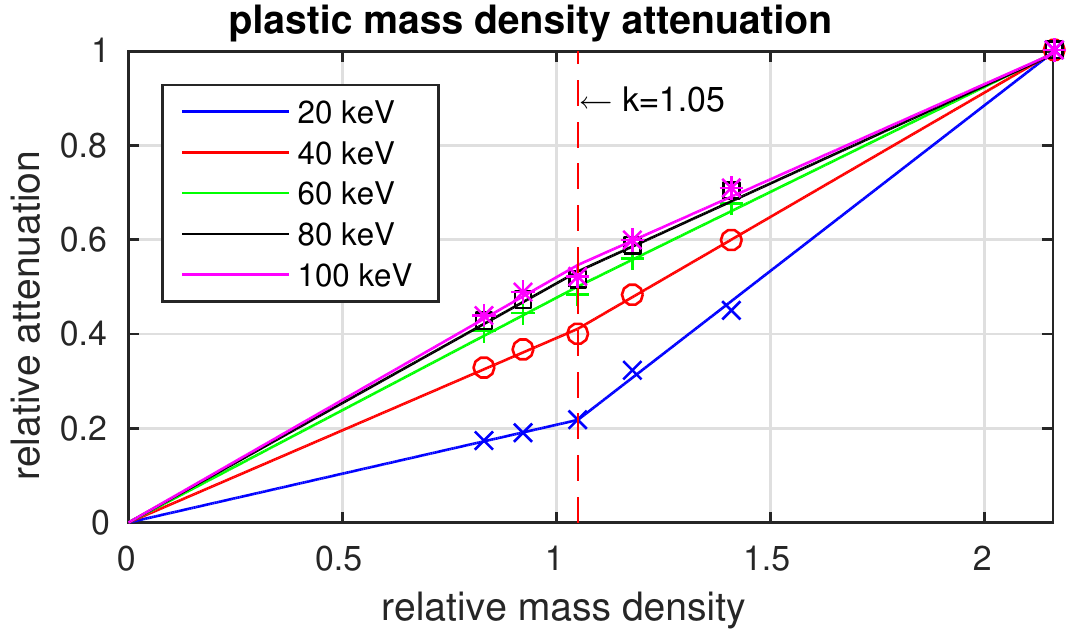}
	\end{subfigure}
	\hfil
	\begin{subfigure}[b]{0.49\textwidth}
		\includegraphics[width=\textwidth]{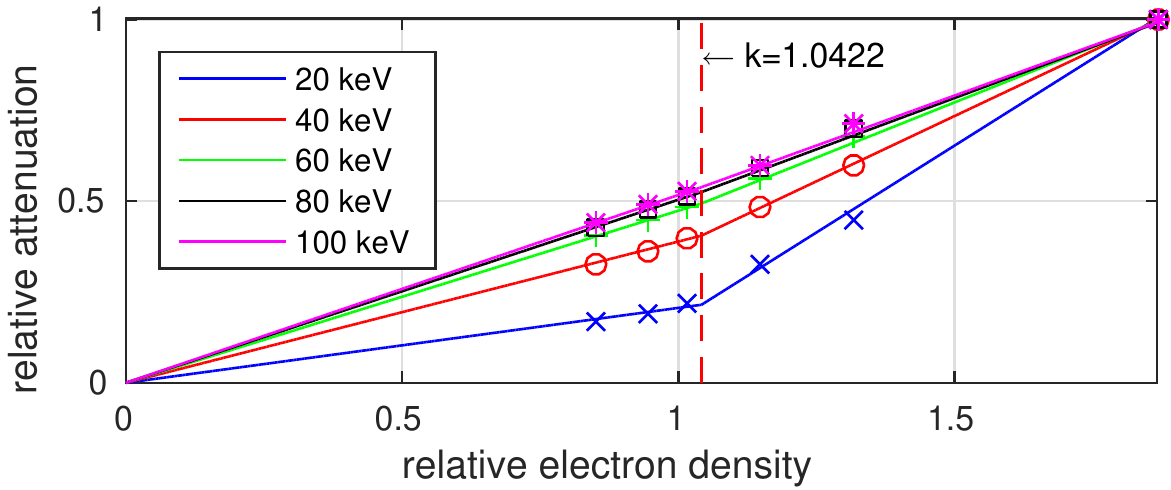}
	\end{subfigure}
	\caption{Relation between mass and electron density for a collection of synthetic materials. As with Figure~\ref{fig:pe_atten}, the attenuation has been normalised to the maximum attenuating material---Teflon in this case---to allow visualisation on a single graph}
	\label{fig:plastic_plot}
\end{figure}

We finally note that our model is still able to account for plastic materials, and the `Polyquant-plastic-$\rho_e$' in Table~\ref{tab:platic_res} is a result of fitting to a family of polymers---we used  Teflon\textsuperscript{\textregistered}, Delrin\textsuperscript{\textregistered}, Plexiglass\textsuperscript{\textregistered}, polystyrene, LDPE (low-density polyethylene) and PMP (polymethylpentene)---and the resulting accuracy is notably very high. We also show the relationship to attenuation for these materials in the graphs in Figure~\ref{fig:plastic_plot}, which confirms that plastics interestingly follow a similar piecewise trend of their own, though this clearly does not extend to Titanium. If one wishes to quantify the attenuation of a mixture of both synthetic and biological tissues, then it seems that no single energy parameterisation would be consistent. One may opt for DECT measurements, but at the cost of requiring two diverse spectral sources.

\subsection{Low Dose Numerical Reconstruction Test} \label{sec:rec_test}
\begin{figure}[!htb]
	\centering
	\begin{subfigure}[b]{0.45\textwidth}
		\includegraphics[width=\textwidth]{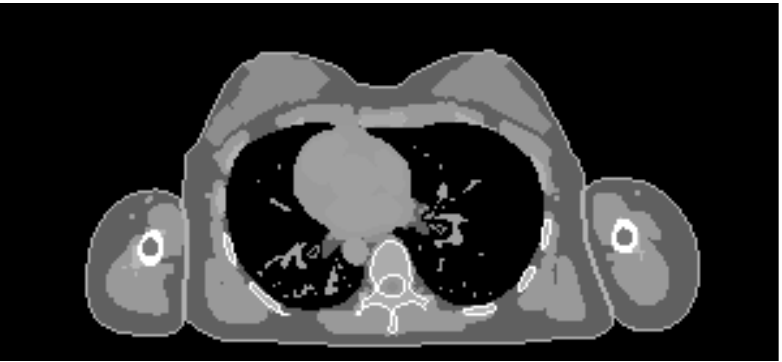}
		\caption{Chest test image}
		\label{subfig:im_chest}
	\end{subfigure}
	\hfil
	\begin{subfigure}[b]{0.45\textwidth}
		\includegraphics[width=\textwidth]{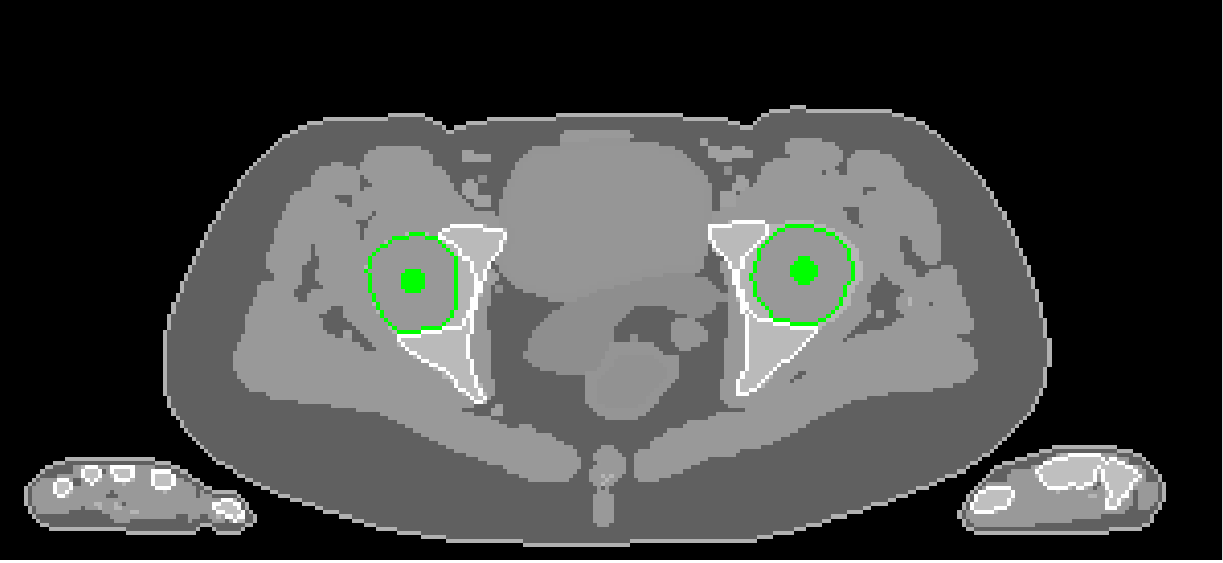}
		\caption{Pelvis test image}
		\label{subfig:im_metal}
	\end{subfigure}
	\caption{Experimental data used: (a) is the oracle chest electron density; and (b) is the oracle pelvis electron density; both have a with display window of [0.8,1.2]}
	\label{fig:data}
\end{figure}

The data we used for our numerical reconstruction test were derived from the Adult Reference Computational Phantom \citep{icrp2009}, which is a segmented image of defined density and chemical composition to represent real tissues. To investigate the ability to image metal implants, we inserted a pair of prosthetic hip joints (marked in green) with a solid titanium pin and shell. The slices through the chest and pelvis---which are the images we selected for testing---are shown in Figure~\ref{fig:data}. The resolution is $299\times137$.

The system geometry used for testing was a flat detector fan-beam CT, which was modelled in the Monte-Carlo package Gate \citep{Jan2011}. We included a focused lead strip collimator to mitigate scatter, a 0.6 mm thick CsI scintillator, a shaped source to put more photons into the centre of the specimen, and generated spectra using SpekCalc \citep{Poludniowski2009c}: 6 mm aluminium filtration at 120 kVp. We simulated a total of $3\times10^9$ photons into 512 detection elements over 360 projections in $1^\circ$ increments, which represented a low dose acquisition. Although the spectrum used to generate the measurements was sampled at $1\,\text{keV}$ increments, for reconstruction we used $N_\xi=21$, which was deemed a sufficient number in \citep{DeMan2001a}, and corresponded to our sampling of mass attenuation coefficients.

\subsubsection{Reconstruction Methods Under Test} \label{sec:rmut}
Most of the methods we tested are iterative reconstructions, with some degree of polyenergetic modelling. For fairness, we used TV regularisation for each \citep{Rudin1992}, and selected the regularisation parameter giving the highest performance in each case --- curves showing the influence of these parameters are shown in Figure~\ref{fig:rmse_curves}. All iterative methods were run for 500 iterations, which resulted in empirical convergence in each case, after which the figure of merit did not changed by more than the reported precision.

Details of the methods under test are:
\begin{itemize}
\item \textit{FBP}: 
Filtered backprojection (FBP) is a popular reconstruction algorithm, that `analytically' approximates the inverse of linearised CT projections \citep{Fessler2014}. In our implementation: we take the logarithm of raw measurements; correct for beam hardening with a water only polynomial fitting \citep{Joseph1978}; apply a ramp filter with Hann windowing, with cut-off frequency optimised to give maximum performance; and finally apply a single backprojection operation $\boldsymbol{\Phi}^T$. To map to density from attenuation coefficient, we use the calibration curve in Figure~\ref{fig:hu_density} and detailed in \cite{Schneider1996}.

We included FBP as a crude baseline and to indicate the level of noise in our system, but we expect it to perform significantly worse than other competitive approaches under test due to the low dose.

\item \textit{PWLS}: 
Penalised weighted-least-squares (PWLS) approximates the CT model as linear by taking the logarithm of the raw measurements \citep{Chang2014}, calibrating them to correct for beam-hardening artefacts for the polyenergetic source \citep{Joseph1978}, and includes a statistical weighting to approximate the Poisson noise in (\ref{equ:poly-poiss}). The objective function is then
\begin{equation} \label{equ:pwls}
\boldsymbol{\mu}_\mathrm{mono} = \argmin_{\boldsymbol{\mu}} (\boldsymbol{\Phi\mu}-\boldsymbol{l})^T\boldsymbol{W}(\boldsymbol{\Phi\mu}-\boldsymbol{l}) + \lambda R(\boldsymbol{\mu}),
\end{equation}
where $\boldsymbol{W}$ is a diagonal statistical weighting matrix with entries $w_{ii}=(y_i-s_i)^2/y_i$, and $\boldsymbol{l}$ is the collection of linearised monoenergetic projections \citep{Chang2014}. Converting $\boldsymbol{\mu}_\mathrm{mono}$ to electron density is then done through a nonlinear calibration according to \citep{Schneider1996}, as with the FBP. It should be noted that we are not actively modelling the metal implant in this case.

\item \textit{Poly-SIR}: 
Polyenergetic statistical iterative reconstruction (Poly-SIR) is the segmented water--bone model from (\ref{equ:water_bone}) \citep{Elbakri2002}. As this requires prior knowledge on material classes, we give it the oracle segmentation of the hard bone structures, and it treats everything else as water. In the pelvis case, we also pass the oracle segmentation of the metal implants, with corresponding mass attenuation coefficients. Since the physical model gives the mass density, we convert to relative electron density where appropriate using the curve shown in Figure~\ref{subfig:den_eden}.

\item \textit{IMPACT}: 
We implement the IMPACT model of \citep{DeMan2001a} as a special case of our generalised fitting in (\ref{equ:de_man}). This allows us to use the same algorithm as presented in Section~\ref{sec:algorithm} for its minimisation. We use a three component piecewise-linear fitting from monoenergetic attenuation at 60 keV to energy dependent attenuation, which accounts for metal implants as well as biological tissues. We then use the same post-processing calibration technique as with FBP and PWLS to convert to either mass or relative electron densities.

\item \textit{Polyquant}: 
In our proposed model, we use the piece-wise linear fitting in Section~\ref{sec:eden_atten} and reconstruction strategy in Section~\ref{sec:quant_recon}. We use Algorithm~\ref{alg1}, but with $10\delta$ which from Figure~\ref{fig:converge} is still rather conservative, but still exhibited a monotonic decrease in the objective function. For the pelvis case, we extend the model with a second `knee' point and linear section to include the attenuation of solid titanium according to the generalisation in (\ref{equ:gen_fit}). We used separate fittings to electron and mass density shown in top and middle graphs in Figure~\ref{fig:pe_atten}, to demonstrate the ability to reconstruct directly into either quantity, where each mapping was a least-squares fitting to the materials in ICRP 89 \citep{icrp2002}.
\end{itemize}

\subsubsection{Quality Assessment Metric}
The metric for quantifying the accuracy of the various methods under test is the root-mean-squared-error (RMSE) in relative electron and mass density, calculated as
\begin{equation}
\sqrt{\frac{1}{N_\mathrm{vox}}\sum_{i=1}^{N_\mathrm{vox}}(\hat{\rho}_i-\rho_i})^2,
\end{equation}
where $\hat{\boldsymbol{\rho}}\in\mathcal{R}^{N_\mathrm{vox}}$ is the estimated mass or electron density of a tested method, and $\boldsymbol{\rho}$ is the ground-truth. As an error, the lower the score represents a higher quantitative accuracy. 

\subsubsection{Reconstruction Results}
\begin{figure}[!htb]
	\centering
	\begin{subfigure}[b]{0.44\textwidth}
		\includegraphics[width=\textwidth]{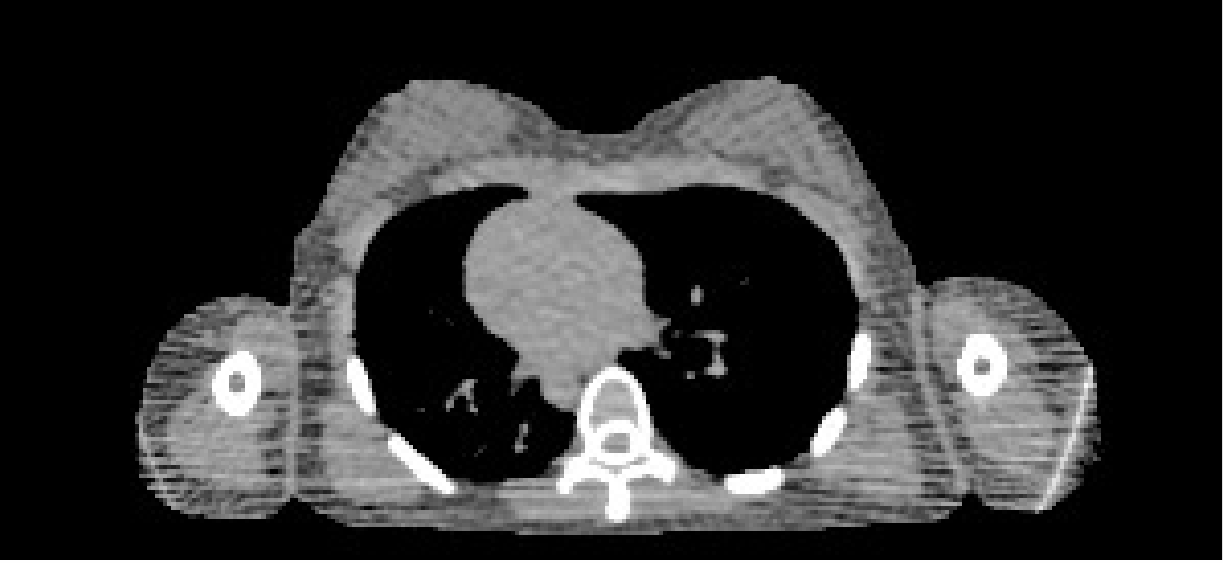}
		\caption{FBP chest}
		\label{subfig:orc_chest}
	\end{subfigure}
	\hfil
	\begin{subfigure}[b]{0.44\textwidth}
		\includegraphics[width=\textwidth]{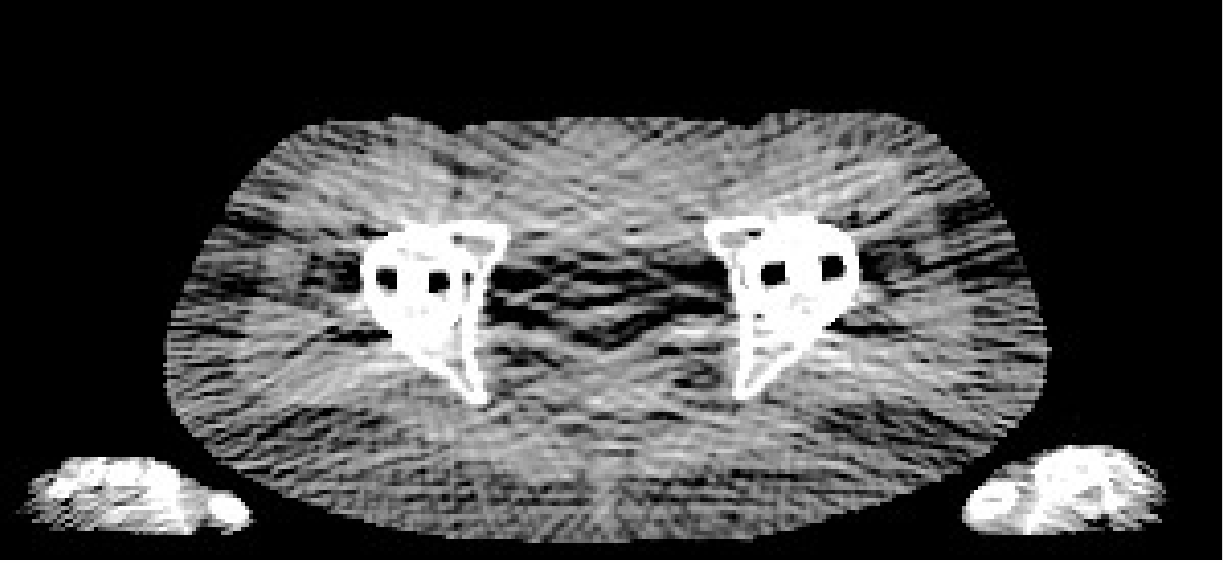}
		\caption{FBP pelvis}
		\label{subfig:orc_metal}
	\end{subfigure}
	\begin{subfigure}[b]{0.44\textwidth}
		\includegraphics[width=\textwidth]{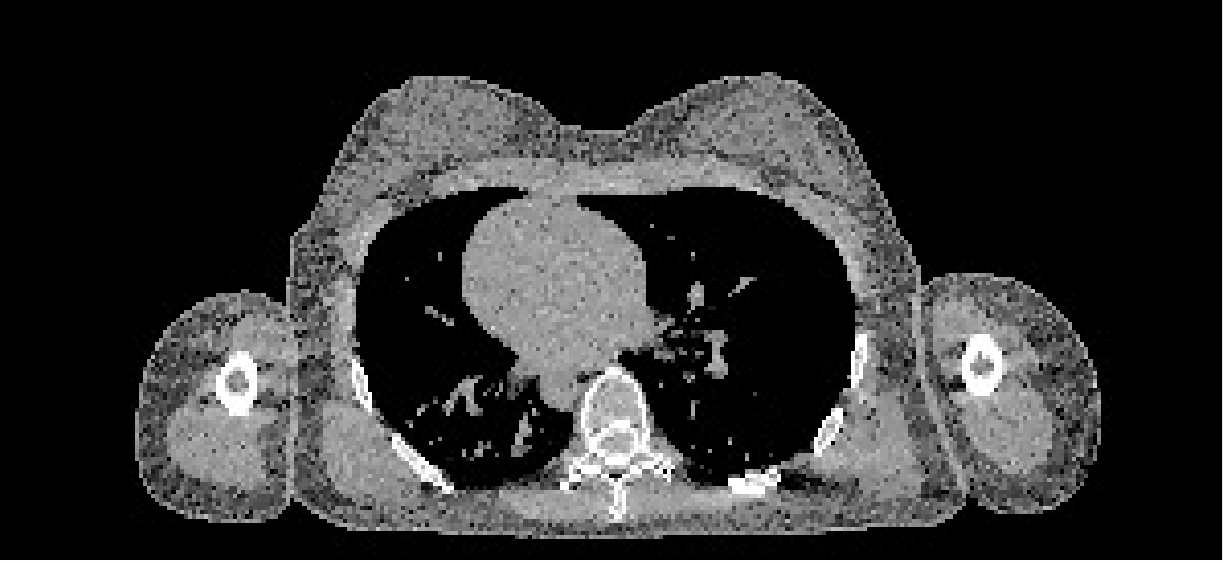}
		\caption{PWLS chest}
		\label{subfig:tv_chest}
	\end{subfigure}
	\hfil
	\begin{subfigure}[b]{0.44\textwidth}
		\includegraphics[width=\textwidth]{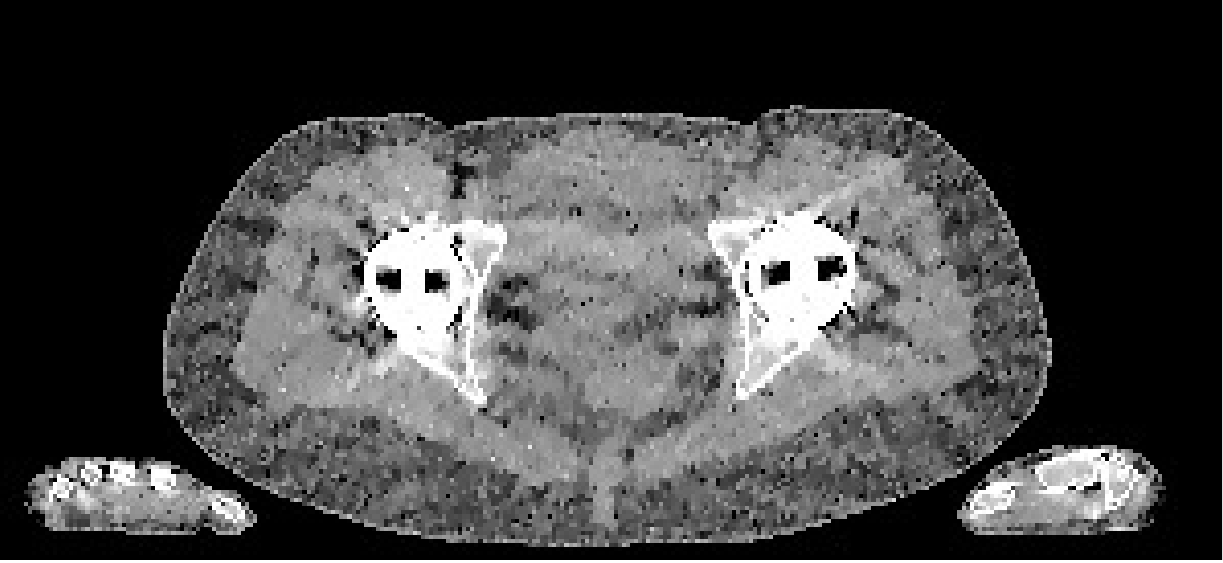}
		\caption{PWLS pelvis}
		\label{subfig:tv_metal}
	\end{subfigure}
	\begin{subfigure}[b]{0.44\textwidth}
		\includegraphics[width=\textwidth]{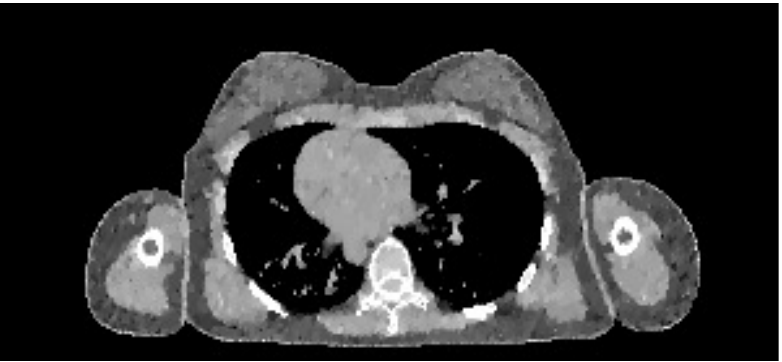}
		\caption{Poly-SIR chest}
		\label{subfig:elekbri_chest}
	\end{subfigure}
	\hfil
	\begin{subfigure}[b]{0.44\textwidth}
		\includegraphics[width=\textwidth]{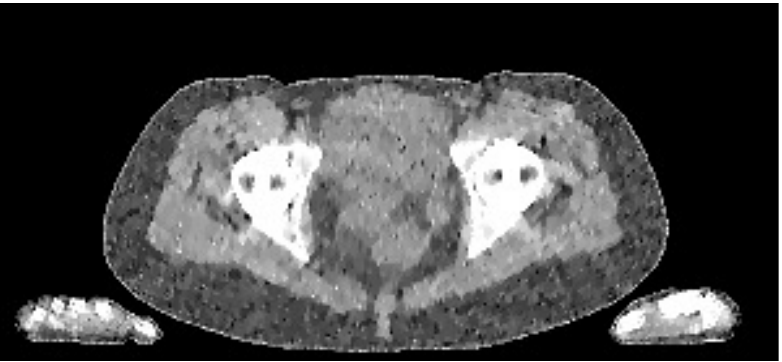}
		\caption{Poly-SIR pelvis}
		\label{subfig:elekbri_metal}
	\end{subfigure}
	\begin{subfigure}[b]{0.44\textwidth}
		\includegraphics[width=\textwidth]{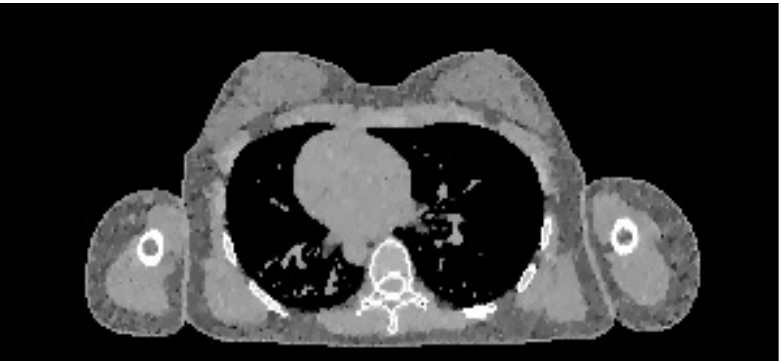}
		\caption{IMPACT chest}
		\label{subfig:de_chest}
	\end{subfigure}
	\hfil
	\begin{subfigure}[b]{0.44\textwidth}
		\includegraphics[width=\textwidth]{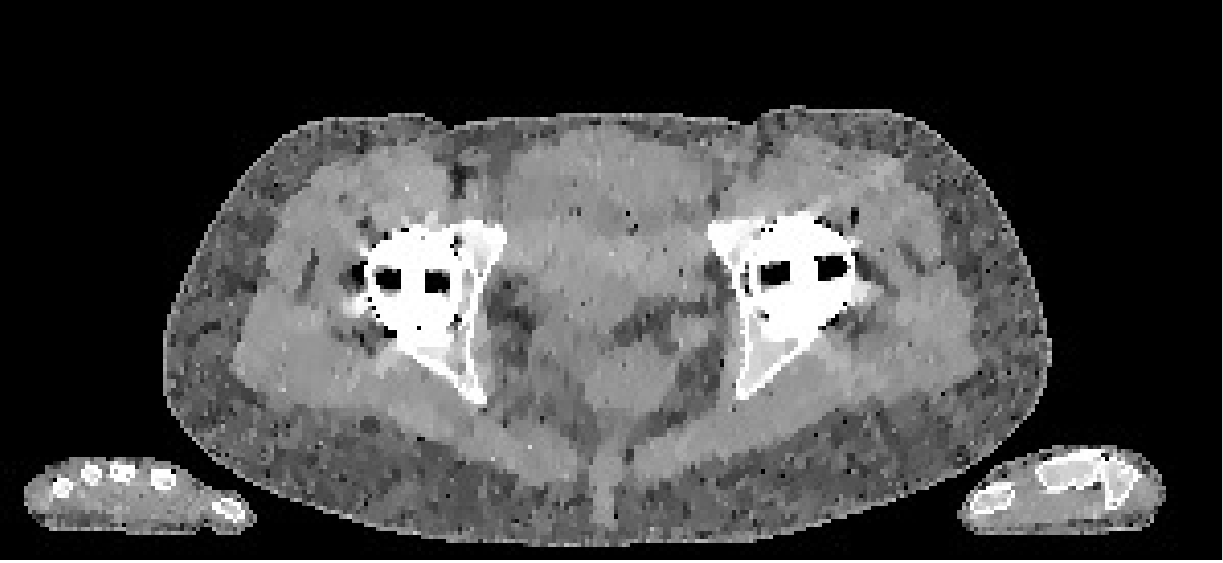}
		\caption{IMPACT pelvis}
		\label{subfig:de_metal}
	\end{subfigure}
	\begin{subfigure}[b]{0.44\textwidth}
		\includegraphics[width=\textwidth]{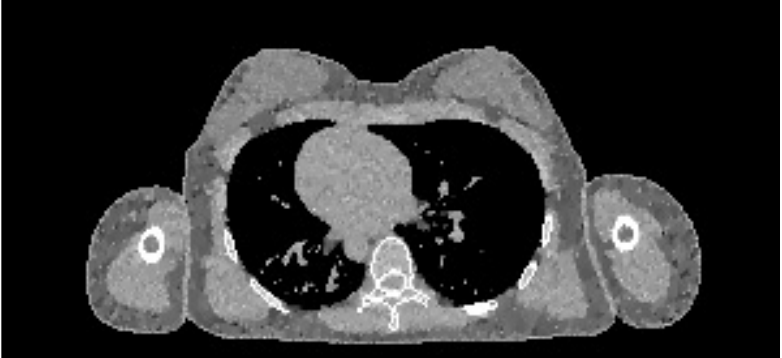}
		\caption{Polyquant chest}
		\label{subfig:eden_chest}
	\end{subfigure}
	\hfil
	\begin{subfigure}[b]{0.44\textwidth}
		\includegraphics[width=\textwidth]{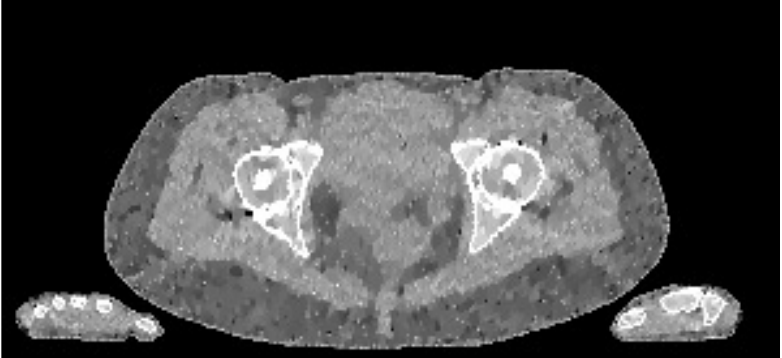}
		\caption{Polyquant pelvis}
		\label{subfig:eden_metal}
	\end{subfigure}
	\caption{Results from electron density reconstruction test for chest and pelvis fan-beam data with display window [0.8,1.2] to aid visualisation of soft tissue and reconstruction artefacts}
	\label{fig:results}
\end{figure}

The results from our reconstruction test are illustrated in Figure~\ref{fig:results} and Table~\ref{tab:results}.
\begin{figure}[!htb]
	\centering
	\begin{subfigure}[b]{0.44\textwidth}
		\includegraphics[width=\textwidth]{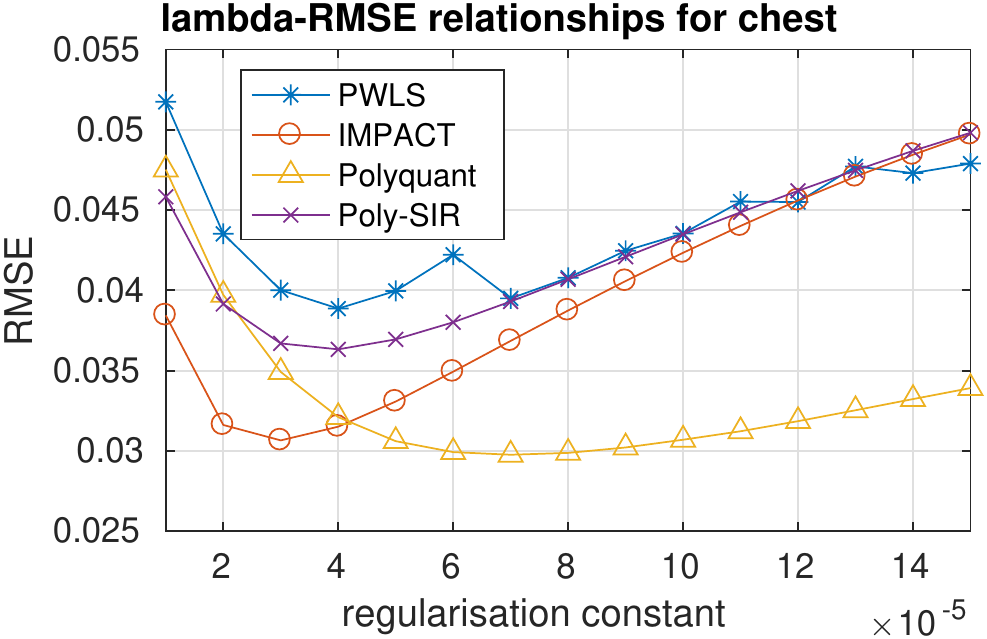}
	\end{subfigure}
	\hfil
	\begin{subfigure}[b]{0.44\textwidth}
		\includegraphics[width=\textwidth]{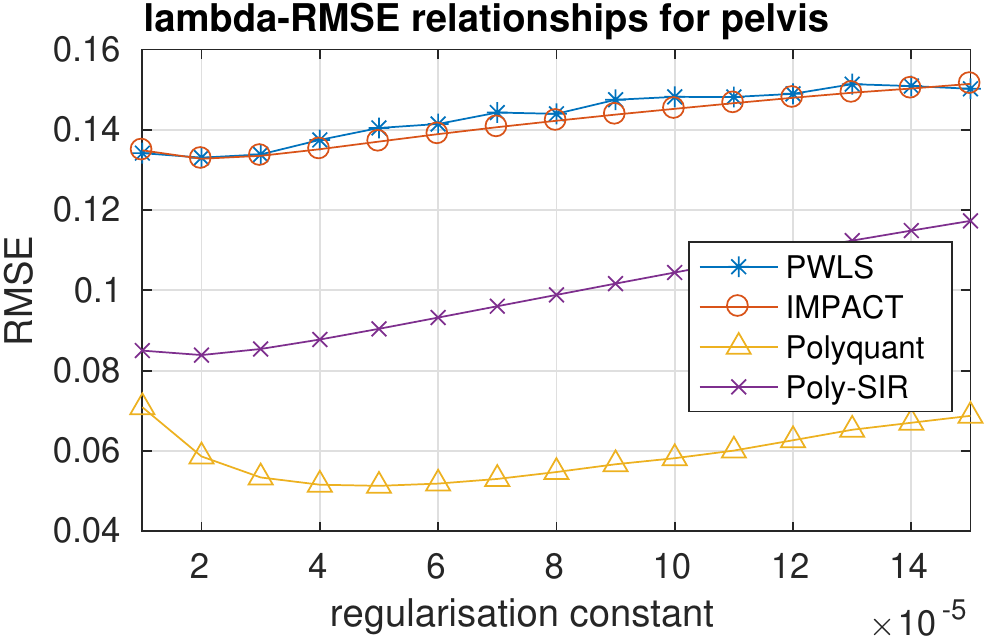}
	\end{subfigure}
	\caption{Curves of electron density RMSE against regularisation parameter for iterative methods for iterative methods under test in chest and pelvis data sets. The minima from each curve is selected in each case given the reconstructions shown in Figure~\ref{fig:results} and quantitatively assessed in Table~\ref{tab:results}}
	\label{fig:rmse_curves}
\end{figure}
\begin{table}[!htb]
	\caption{Quantitative results: relative electron density ($\rho_e$) and mass density ($\rho$) RMSE}
	\label{tab:results}
	\centering
	$
	\begin{array}{c|c|c|c|c|}
	\cline{2-5}
	&\multicolumn{2}{c}{\text{chest data}}&\multicolumn{2}{|c|}{\text{pelvis data}}\\
	\cline{2-5}
	\text{Scheme} & \rho_e \text{ RMSE} & \rho \text{ RMSE} & \rho_e \text{ RMSE} & \rho \text{ RMSE} \\
	\hline
	\text{FBP} & 0.136 & 0.140 & 0.201 & 0.226 \\
	\text{PWLS} & 0.0389 & 0.0426 & 0.133 & 0.166 \\
	\text{Poly-SIR} & 0.0363 & 0.0385 & 0.0839 & 0.101 \\
	\text{IMPACT} & 0.0307 & 0.0326  & 0.133 & 0.159 \\
	\text{Polyquant} & \mathbf{0.0298} & \mathbf{0.0316} & \mathbf{0.0513} & \mathbf{0.0746} \\
	\hline
	\end{array}
	$
\end{table}

In terms of both electron density and mass density accuracy, our proposed method is the best performing method under test. The lack in estimation accuracy in using the water--bone model with Poly-SIR is likely to be due to discrepancies in adipose and spongy bone tissues from water, which were illustrated in Figure~\ref{fig:tissue_fit}. We note that Poly-SIR is second best performing in the pelvis case, but it was provided with oracle information on the implant and hard bone, which would be difficult to segment in practice from a preliminary reconstruction such as FBP due to its high noise. IMPACT is understandably very closely performing to our method in the chest case, as these performed similarly in the model test, but the advantage of fitting at each energy is clear in the case of the metal implant, where the IMPACT performance is similar to that of PWLS.

The relationship between regularisation strength and quantitative performance of the iterative methods are shown in Figure~\ref{fig:rmse_curves}. In both cases, this demonstrates that the Polyquant not only reaches the best performance, especially in the pelvis case, but its numerical accuracy is reasonably robust to the setting of this parameter. The difference in the location of the minima between methods may be accounted for by the different scales in image parameter --- for example IMPACT uses the monoenergetic attenuation at 60 keV, whereas Polyquant uses the physical density, and these have a relative difference in intensities. Another implication from these curves is that although the Polyquant chest image in Figure~\ref{subfig:eden_chest} exhibits a higher level of intratissue variation than IMPACT, this will be mitigated by increasing the regularisation parameter, which can increase by over 40\% whilst maintaining the best numerical performance.

Another interesting feature of the results is the difference between electron and mass density scores. Although it may seem counter-intuitive that the Poly-SIR for example would have a higher error before calibration than after, this is due to the RMSE being absolute and not relative. Since from Figures~\ref{subfig:den_eden} and \ref{fig:pe_atten} the mass densities of materials are on average higher than relative electron density, this will account for the difference.

\subsection{Cone-Beam CT Validation} \label{sec:ccbt_recon}
To validate our method, we also tested its ability to perform quantitative reconstruction from real X-ray cone-beam CT (CBCT) measurements. For this, we acquired a scan with a Varian\textsuperscript{\textregistered} TrueBeam\texttrademark\ On-Board Imager\textsuperscript{\textregistered} of a CIRS STEEV head phantom. The phantom consists of synthetic resins to mimic the attenuating properties of human tissues, allowing quantitative assessment of relative electron density accuracy. There was also a metal structure in the centre of the phantom, consisting of the plug section from a PTW PinPoint\textsuperscript{\textregistered} ionisation chamber, allowing us to investigate the mitigation of metal induced artefacts.

\subsubsection{CBCT Data Processing}
Our CBCT acquisition consists of 499 projections at a 100 kVp tube potential and 20 mA current for 15 mSec on each, which were the default settings for a head acquisition. Compared to the numerical test, the relative X-ray flux was roughly $2.5\times$ higher, which coupled with the smaller specimen volume and larger number of projections, implies that this test was at a significantly higher dose. 

The raw measurements were all pre-corrected for detector responses, and the effect of bow-tie shifting with gantry rotation, with default TrueBeam\texttrademark\ corrections to give $\boldsymbol{y}$ in (\ref{equ:poly-poiss}). The scatter estimate $\boldsymbol{s}$ in (\ref{equ:poly-poiss}) was also taken from the Varian system's default scatter correction. For testing the FBP---realised with the Feldkamp--Davis--Kress method \citep{Feldkamp1984}---and PWLS, we calculated the linearised projection vector $\boldsymbol{l}$ in (\ref{equ:pwls}).

In the case of the fully polyenergetic reconstruction methods Poly-SIR, IMPACT and Polyquant, we require explicit knowledge of the X-ray spectrum $b_i(\xi)$ in (\ref{equ:poly-poiss}). Due to the variable thickness of Aluminium in the bow-tie filter, this will be spatially varying, and we calculated it analytically from the appropriate spectrum in the system's calibration parameters, and the spectral response of the various metal filters and scintillator in the beam path. As in our other experimental sections, we discretised the spectrum into 21 energies, and used the same parameters as were fitted in Section~\ref{sec:model_test} for the ICRP biological tissues, but supplemented by the mass attenuation of the metal implant according to information provided by the manufacturer.

\subsubsection{CBCT Reconstruction}
\begin{figure}[!htb]
	\centering
	\begin{subfigure}[b]{0.44\textwidth}
		\includegraphics[trim=4cm 2.5cm 4cm 3.5cm,clip=true,width=\textwidth,angle=180]{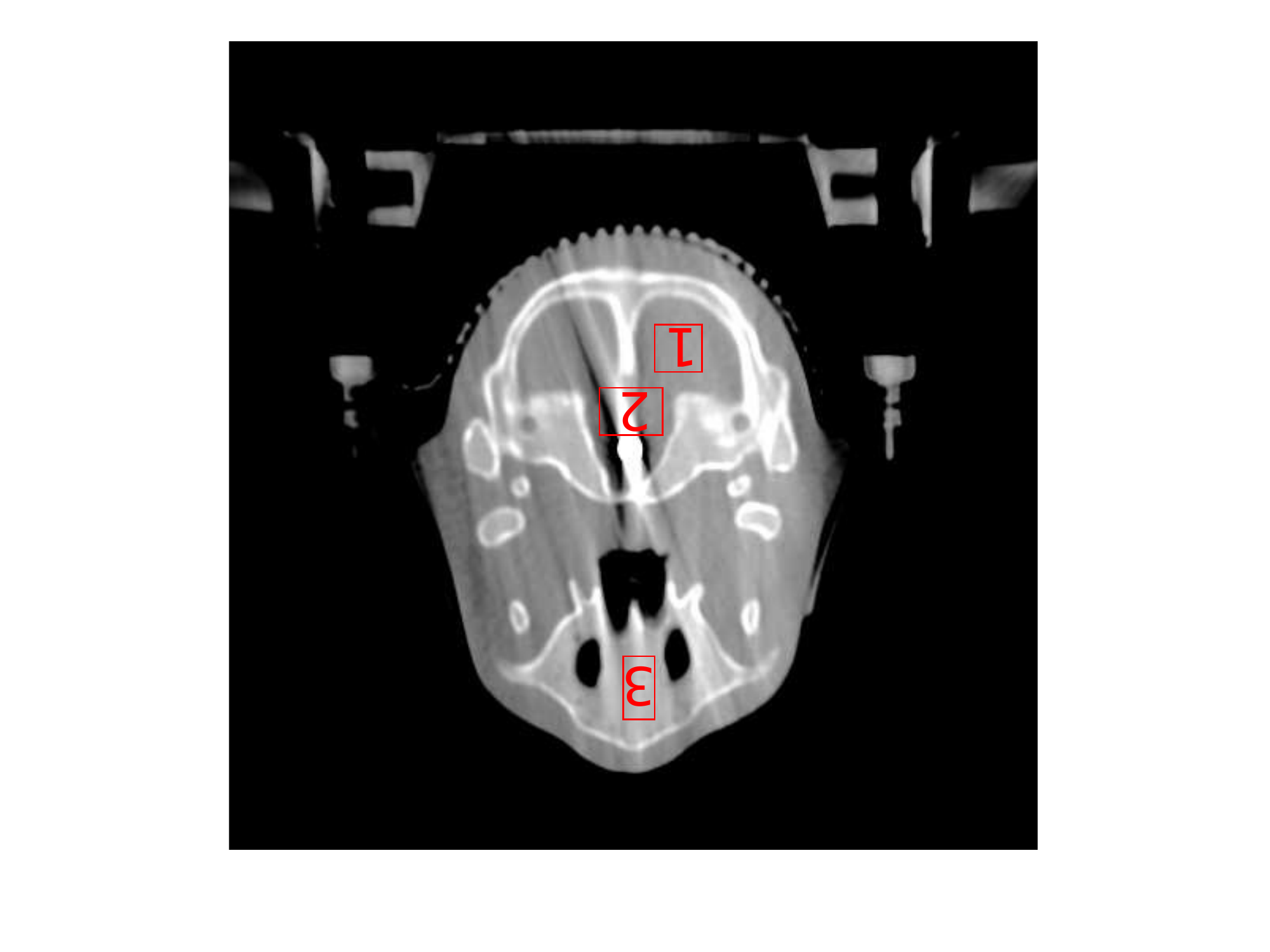}
		\caption{FBP CBCT}
		\label{subfig:fbp_sk200}
	\end{subfigure}
	\hfil
	\begin{subfigure}[b]{0.44\textwidth}
		\includegraphics[trim=4cm 2.5cm 4cm 3.5cm,clip=true,width=\textwidth,angle=180]{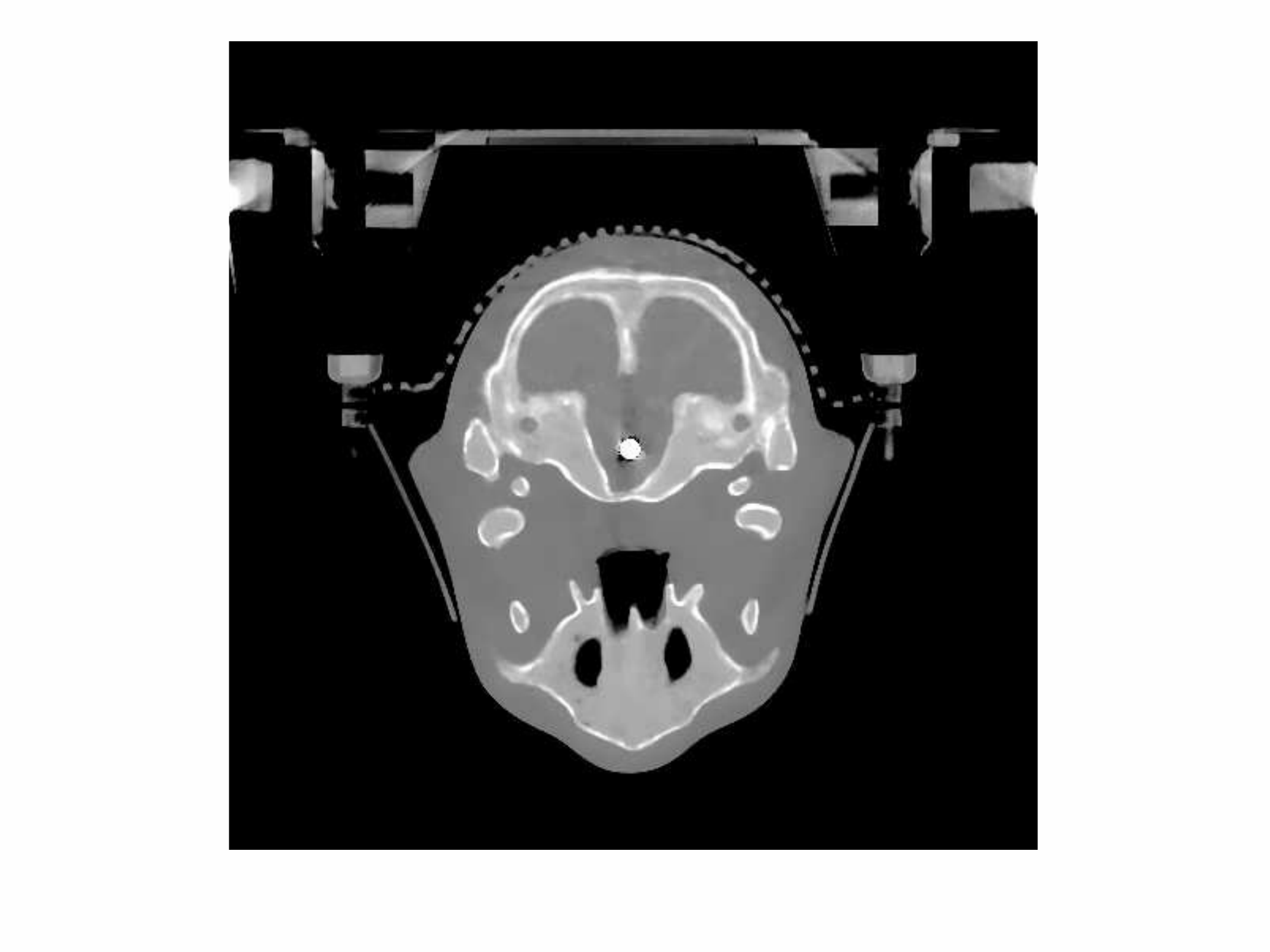}
		\caption{PWLS CBCT}
		\label{subfig:pwls_sk200}
	\end{subfigure}
	\begin{subfigure}[b]{0.44\textwidth}
		\includegraphics[trim=4cm 2.5cm 4cm 3.5cm,clip=true,width=\textwidth,angle=180]{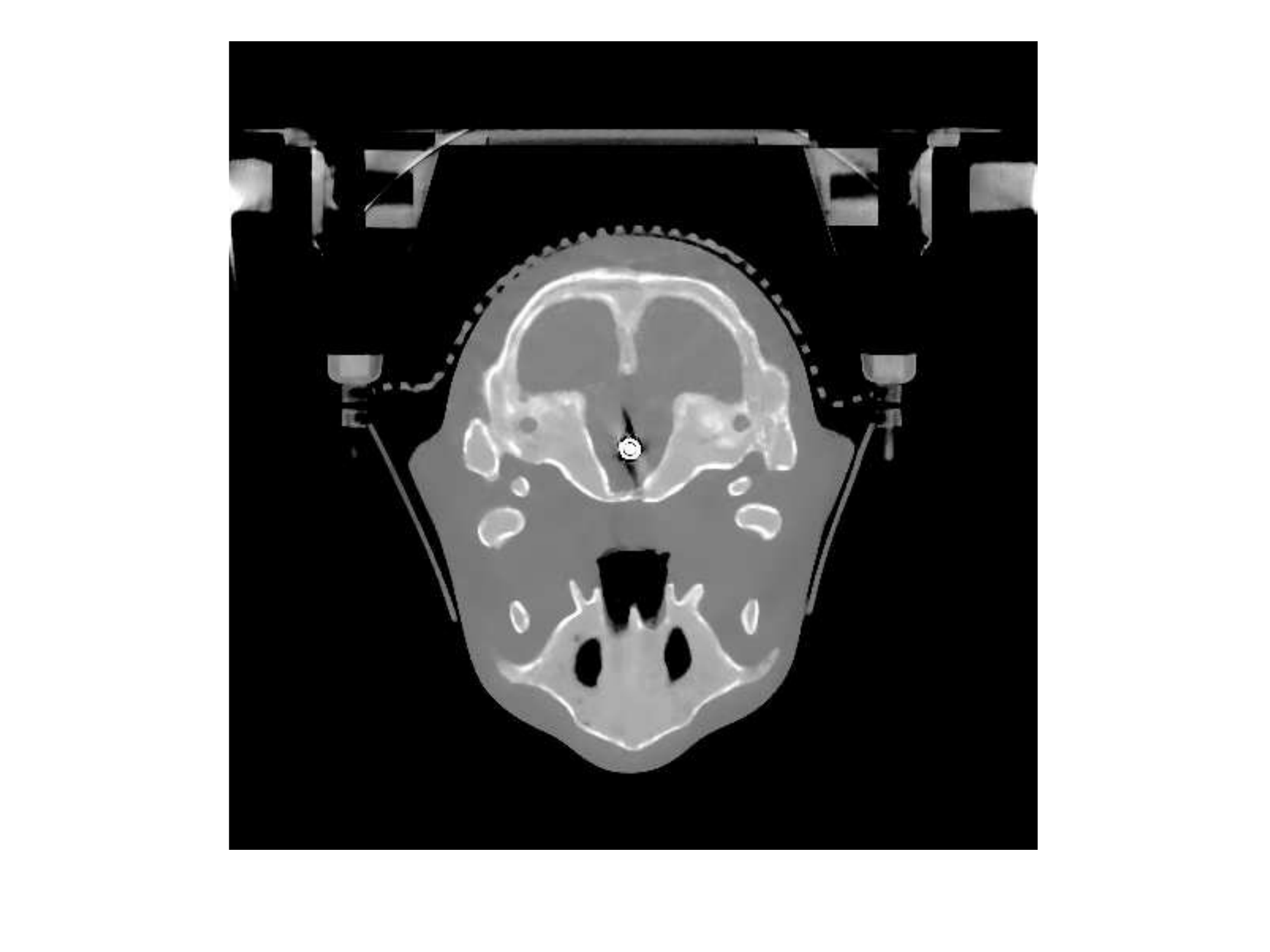}
		\caption{Poly-SIR CBCT 1}
		\label{subfig:elekbri1_sk200}
	\end{subfigure}
	\hfil
	\begin{subfigure}[b]{0.44\textwidth}
		\includegraphics[trim=4cm 2.5cm 4cm 3.5cm,clip=true,width=\textwidth,angle=180]{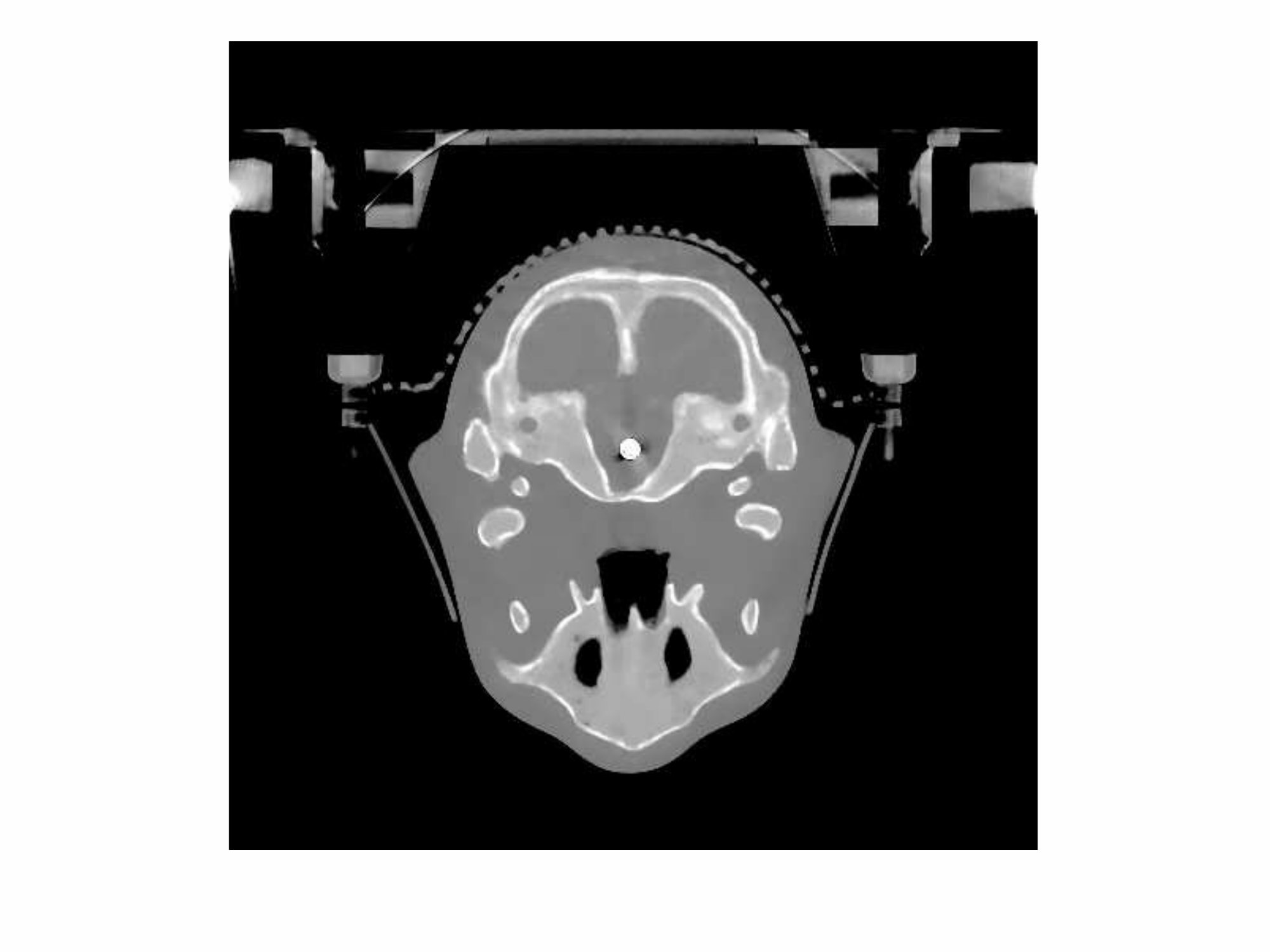}
		\caption{Poly-SIR CBCT 2}
		\label{subfig:elekbri2_sk200}
	\end{subfigure}
	
	\begin{subfigure}[b]{0.44\textwidth}
		\includegraphics[trim=4cm 2.5cm 4cm 3.5cm,clip=true,width=\textwidth,angle=180]{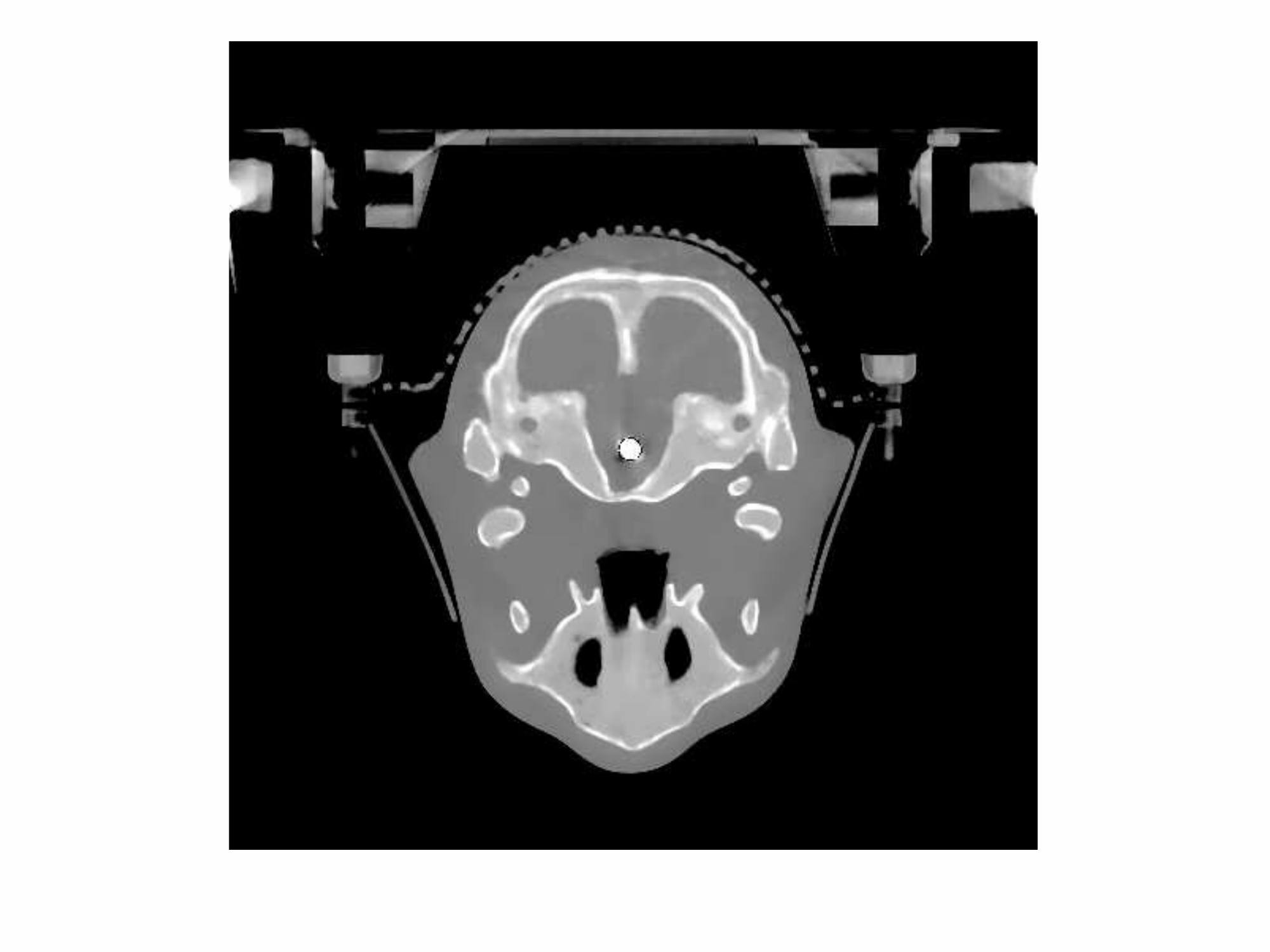}
		\caption{IMPACT CBCT}
		\label{subfig:polyquant_sk200}
	\end{subfigure}
	\hfil
	\begin{subfigure}[b]{0.44\textwidth}
		\includegraphics[trim=4cm 2.5cm 4cm 3.5cm,clip=true,width=\textwidth,angle=180]{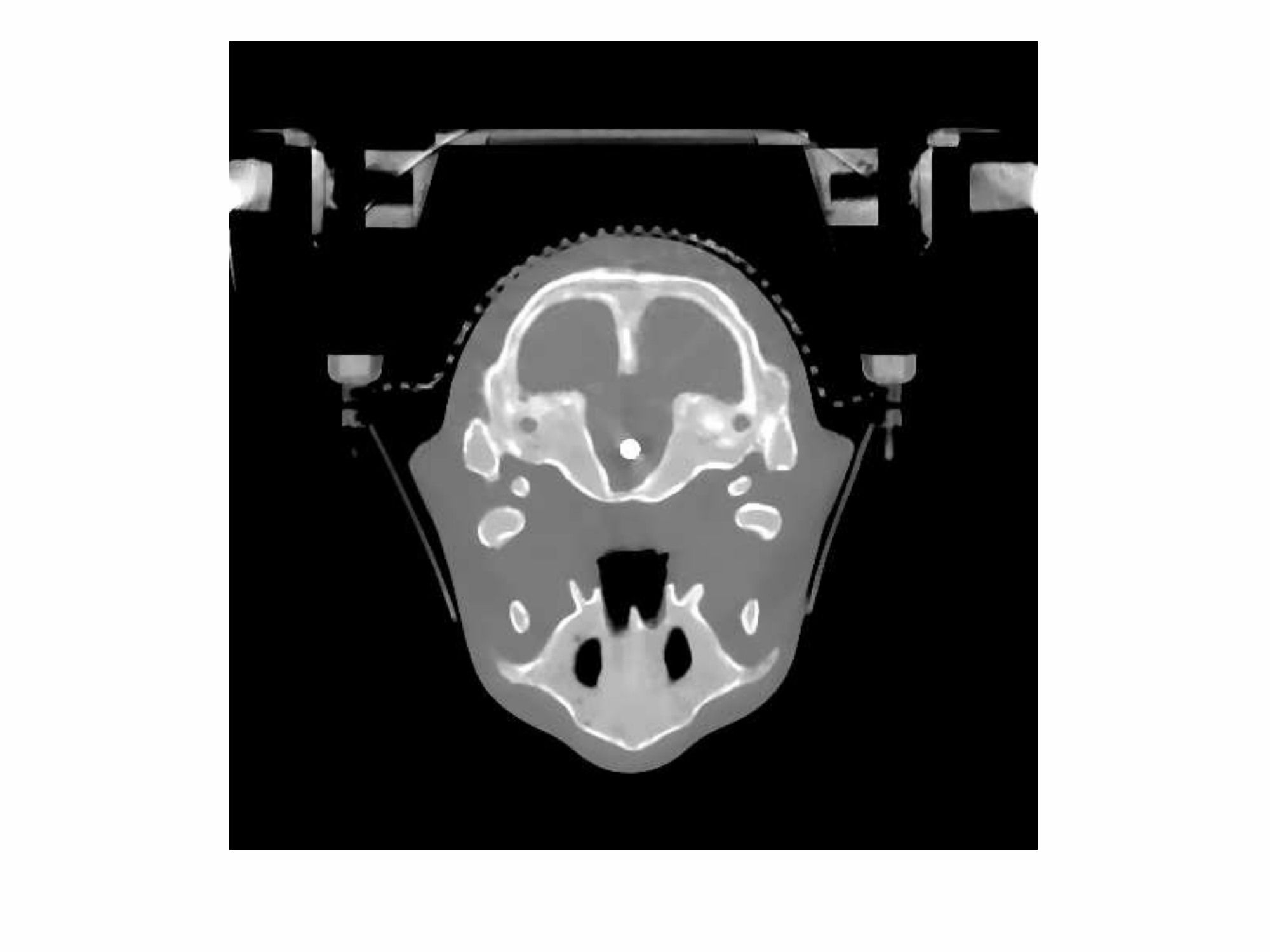}
		\caption{Polyquant CBCT}
		\label{subfig:elekbri_sk200}
	\end{subfigure}
	
	\caption{Results from electron density reconstruction from real CBCT data showing slice 83, where each is shown with display window [0.7,1.4]: (c) is Poly-SIR given a bone and metal segmentation derived from the FBP; (d) uses a segmentation derived from the PWLS}
	\label{fig:cbct_recon}
\end{figure}

For reconstruction, we mapped into a resolution of $512\times 512 \times 144$, and used each method as detailed in Section~\ref{sec:rmut}. We ran each iterative method for 500 iterations. For the regularisation parameter, we heuristically used $0.5\lambda_\textrm{pelvis}$, where $\lambda_\textrm{pelvis}$ were the same TV regularisation parameters from the digital pelvis experiment, and gave good empirical performance on the CBCT data. Finally, for the bone and metal segmentations required for \textit{Poly-SIR}, we obtained these through applying thresholds on the FBP and PWLS separately. To illustrate the critical role of this step, we have shown both images in Figures~\ref{subfig:elekbri1_sk200} and \ref{subfig:elekbri2_sk200}.

\begin{figure}[!htb]
	\centering
	\begin{subfigure}[b]{0.18\textwidth}
		\includegraphics[width=\textwidth,angle=180]{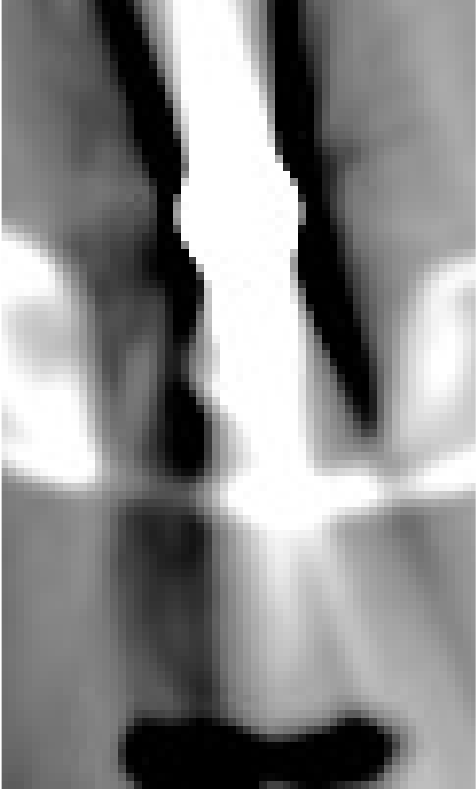}
		\caption{FBP metal}
	\end{subfigure}
	\begin{subfigure}[b]{0.18\textwidth}
		\includegraphics[width=\textwidth,angle=180]{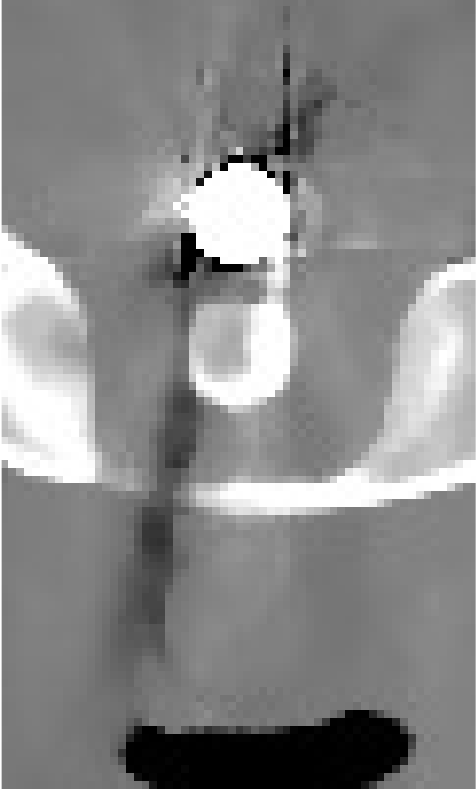}
		\caption{PWLS metal}
	\end{subfigure}
	\begin{subfigure}[b]{0.18\textwidth}
		\includegraphics[width=\textwidth,angle=180]{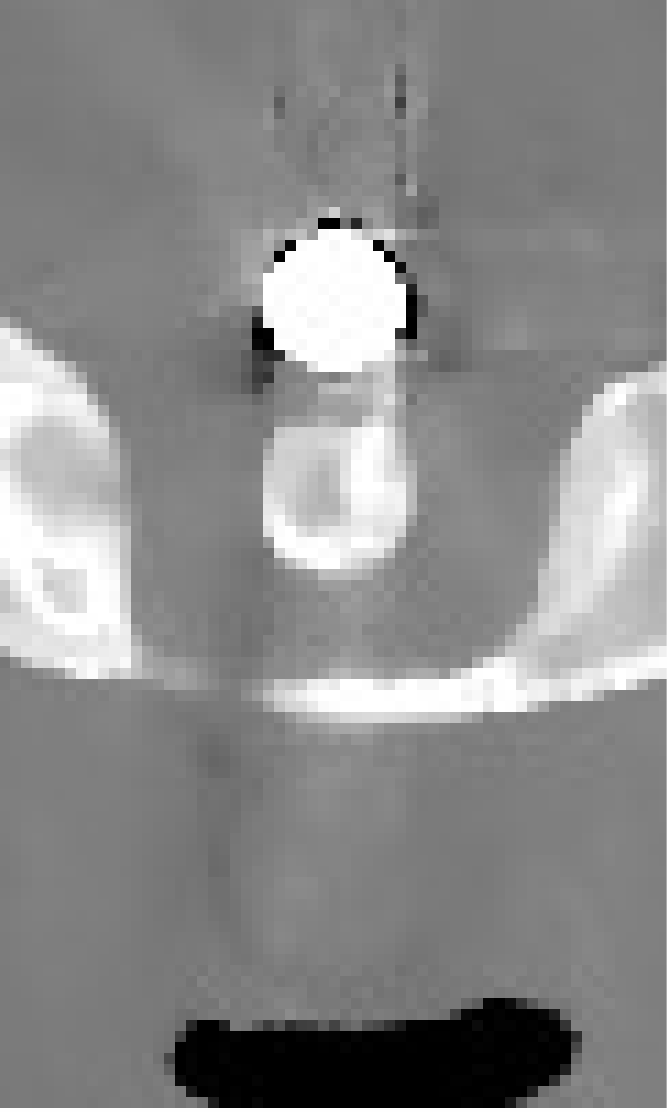}
		\caption{Poly-SIR metal}
		\label{subfig:polysir_metal}
	\end{subfigure}
	\begin{subfigure}[b]{0.18\textwidth}
		\includegraphics[width=\textwidth,angle=180]{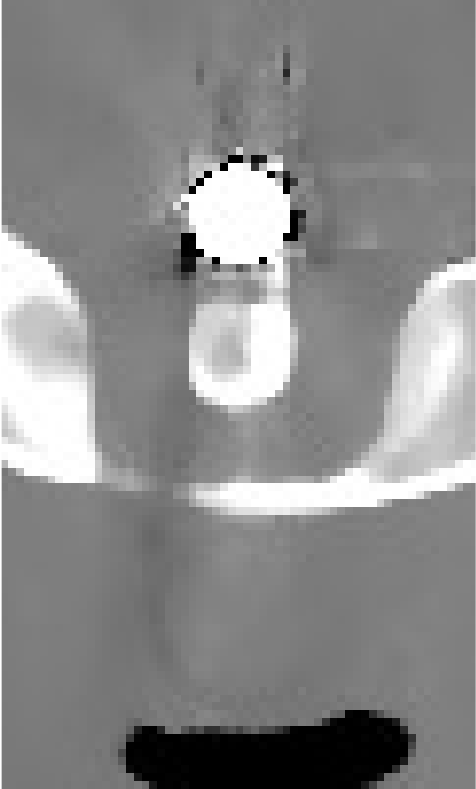}
		\caption{IMPACT metal}
	\end{subfigure}
	\begin{subfigure}[b]{0.18\textwidth}
		\includegraphics[width=\textwidth,angle=180]{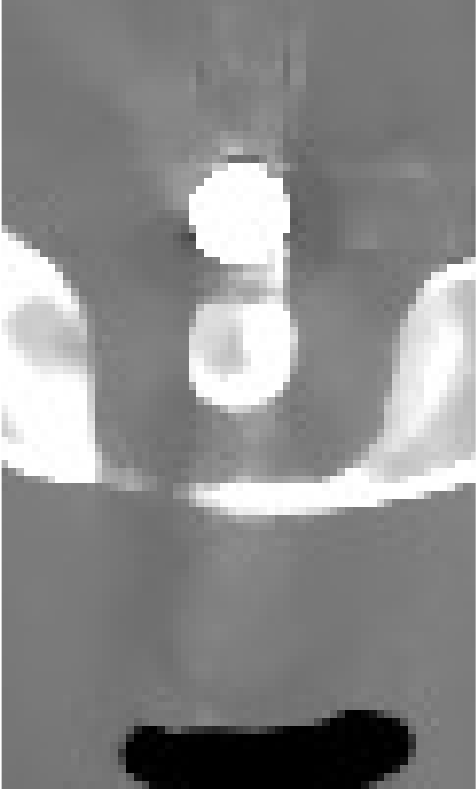}
		\caption{Polyquant metal}
	\end{subfigure}

	\caption{Metal insert visualisation of slice 91 from CBCT reconstructions with display window [0.7,1.4]. The Poly-SIR is the version based upon segmentation from the PWLS as that in Figure~\ref{subfig:elekbri1_sk200}}
	\label{fig:insert_recon}
\end{figure}

Reconstructions of the 83\textsuperscript{rd} slice from the CBCT data are shown in Figure~\ref{fig:cbct_recon}, along with a region from the 91\textsuperscript{st} slice containing a higher mass of metal in Figure~\ref{fig:insert_recon}. Although the FBP appears to suffer strongly from the presence of the metal structure, most of the iterative methods mitigate its effect considerably, with the PWLS showing a more pronounce dark region in the soft tissue. The Poly-SIR based on a segmentation from the FBP does suffer from its streaking as shown in Figure~\ref{subfig:elekbri2_sk200}. Although a better performance is achieved through segmenting from the PWLS as in Figure~\ref{subfig:elekbri2_sk200}, this will have a considerably higher total computational cost; the Poly-SIR region in Figure~\ref{subfig:polysir_metal} is based on this PWLS initialisation. The preservation of bone structure appears to be similar between all the iterative methods. The Polyquant method appears to have the smoothest regions around the metal implant, whilst maintaining the same level of structure in the bone and objects surrounding the phantom.

%
%

To evaluate the quantitative accuracy of each reconstruction, we isolated regions of interest (ROI) in the 83\textsuperscript{rd} slice located in soft tissue and spongy bone --- these are shown in Figure~\ref{subfig:fbp_sk200}. The RMSE of these regions are calculated relative to the electron density of the soft tissue and bone equivalent resins, and shown in Table~\ref{tab:cbct_results}.
\begin{table}[!htb]
	\caption{Quantitative CBCT results: RMSE of relative electron density in regions shown in Figure~\ref{subfig:fbp_sk200}}
	\label{tab:cbct_results}
	\centering
	$
	\begin{array}{c|c|c|c|c|c}
	\text{RMSE} & \text{FBP} & \text{PWLS} & \text{Poly-SIR} & \text{IMPACT} & \text{Polyquant}\\
	\hline
	\text{ROI 1} & 0.0583 & 0.0170 & 0.0200 & 0.0154 & \mathbf{0.0100}\\
	\text{ROI 2} & 0.231 & 0.0235 & 0.0256 & 0.0199 & \mathbf{0.0166}\\
	\text{ROI 3} & 0.0342 & 0.0376 & 0.0333 & 0.0178 & \mathbf{0.0128}\\
	\hline
	\end{array}
	$
\end{table}

From Table~\ref{tab:cbct_results}, we note that our proposed model is the most accurate method under test by at least 16\% over other approaches in all regions.

\section{Discussion} \label{sec:discuss}
Aspects of our method we have not evaluated in this study are its robustness, and the practicality of its computational implementation, though these are both worth discussing. In the first case, we have noted that due to the discontinuous gradient from the piecewise linear fitting, there is no theoretical guarantees for convergence. Two approaches we have adopted in implementations are using a smooth function for $f(x)$ in (\ref{equ:logit-fit}) such as a generalised logistic, and connecting the two linear fits with a quadratic function for some interval around the `knee'. However both options increase the computational cost of the gradient term in (\ref{equ:deriv}) considerably, and we have found neither give any empirical advantage in convergence or accuracy over just using the non-smooth version. Since it is common in CT reconstruction to use empirically well performing methods that have no convergence guarantees such as pre-computed curvatures in separable paraboloid surrogates \citep{Erdogan1999} or ordered subsets \citep{Search1999a}, we believe this is reasonable. We also highlight, that from Figure~\ref{fig:converge}, our method does empirically appear to converge even for very aggressive step size multiplication factors.

Another potential robustness issue with any polyenergetic model are partial volume effects, where a discretised voxel contains different classes of material. Due to its linearity, given that the two materials belong to the same fit interval, such as lung and fat or muscle and bone as in Figure~\ref{fig:pe_atten}, then our model would correctly estimate the attenuation from these materials. If a voxel contains materials from different intervals however, such as fat and muscle or metal and bone, then our model will overestimate the attenuation, according to the trends in Figure~\ref{fig:pe_atten}. Although we have not evaluated the degree of this effect, we note it will be in common with other approaches \citep{Elbakri2002,Elbakri2003,DeMan2001a}, and may be mitigated by increasing the resolution of reconstruction.

The performance of iterative methods under a low dose acquisition is an important consideration for reducing the amount of ionising radiation delivered to the patient. By performing our numerical test at a low dose, we demonstrated the method is relatively robust in this setting. Between the iterative methods tested, PWLS is expected to suffer the most from a low dose, due to its approximation to the noise model, and linearisation of the projections that become unstable for very low photon fluxes \citep{Chang2014}, and has been shown to perform worse at low doses than `pre-log' methods such as Polyquant, Poly-SIR and IMPACT \citep{Fu2014}. Between these three method, whether there exists a relative change of interplay between partial volume effect photon flux and accuracy is yet to be determined, although Polyquant is likely to benefit from its superior attenuation modelling.

In terms of computational cost, our method is comparable to other full gradient iterative methods. Given that the bottleneck is in calculating the forward- and back-projection operators $\boldsymbol{\Phi}$ and $\boldsymbol{\Phi}^T$, then we note that each gradient step in (\ref{equ:deriv}) may be implemented with 3 forward and 2 backward evaluations, which represents a $2.5\times$ larger cost than PWLS, which is the same as our implementation of IMPACT. Comparatively, Poly-SIR has a $2\times$ larger cost than PWLS, which given the consistent accuracy advantage of our model is unlikely to be worth this slight speed advantage. For faster implementation, we suggest that algorithmic acceleration such as ordered subsets \citep{Search1999a,Wang2015} and using parallel hardware \citep{Yan2008} are likely suitable, and will be investigated in future work.

\section{Conclusions} \label{sec:conclusions}
We have introduced a general quantitative attenuation model, which allows direct inference of mass or electron density from raw CT measurements with a single polyenergetic source. Not only have we demonstrated this allows more accurate modelling than explicit physical models such as water--bone or photoelectric--Compton, but have shown how it may be exploited in a flexible reconstruction algorithm that allows accurate quantitative medical imaging, even with metal implants and real CBCT data. As with other single source methods, we have highlighted its inconsistency between synthetic and biological tissues, but this may not be of relevance for medical imaging, in which opting for the more general DECT model is significantly less accurate over materials of interest. Since our method has a similar computational cost to other iterative approaches, but offers markedly higher accuracy, it offers both a practical and beneficial approach to CT imaging.

\section*{Acknowledgements}
The authors would like to sincerely thank Dr Adam Wang from Varian Medical Systems, for preprocessing our cone-beam CT data and providing invaluable information and advice about the imaging system. This work was supported by the Maxwell Advanced Technology Fund, EPSRC DTP studentship funds and ERC project: C-SENSE (ERC-ADG-2015-694888). MD is also supported by a Royal Society Wolfson Research Merit Award.

\bibliographystyle{apalike}
\bibliography{/Users/jon/Documents/Bibtex/quant.bib}
\end{document}